\documentclass[traditabstract]{aa}

\usepackage{graphicx}
\usepackage{txfonts}
\usepackage{natbib}
\usepackage{lscape}
\usepackage{longtable}

\begin{document}

\def\deg{$^{\circ}$}
\newcommand{\eg}{{\it e.g.}}
\newcommand{\ie}{{\it i.e.}}
\newcommand{\minusone}{$^{-1}$}
\newcommand{\kms}{km~s$^{-1}$}
\newcommand{\kmsm}{km~s$^{-1}$~Mpc$^{-1}$}
\newcommand{\Ha}{$\rm H\alpha$}
\newcommand{\Hb}{$\rm H\beta$}
\newcommand{\hi}{{H{\sc i}}}
\newcommand{\hii}{{H{\sc ii}}}
\newcommand{\nii}{\ion{N}{2}}
\newcommand{\rband}{{\em r}-band}
\newcommand{\iband}{{\em I}-band}
\newcommand{\zband}{{\em z}-band}
\newcommand{\rd}{$r_{\rm d}$}
\newcommand{\whi}{$W_{50}$}
\newcommand{\ds}{$\Delta s$}
\newcommand{\x}{$\times$}
\newcommand{\about}{$\sim$}
\newcommand{\Msun}{M$_\odot$}
\newcommand{\Lsun}{L$_\odot$}
\newcommand{\Mhi}{$M_{\rm HI}$}
\newcommand{\Mst}{$M_\star$}
\newcommand{\must}{$\mu_\star$}
\newcommand{\nuvr}{NUV$-r$}
\newcommand{\Rinz}{$R_{50,z}$}
\newcommand{\Ropt}{$R_{\rm opt}$}
\newcommand{\sov}{$S_{0.5}$}
\newcommand{\vrot}{$V_{\rm rot}$}
\newcommand{\vs}{$V_{\rm rot}/\sigma$}
\newcommand{\cindx}{$R_{90}/R_{50}$}
\newcommand{\rhalf}{$R_{50}$}
\newcommand{\tmax}{$T_{\rm max}$}

\title{The GALEX Arecibo SDSS Survey}
\subtitle{VI. Second Data Release and Updated Gas Fraction Scaling Relations}

\author
{Barbara Catinella\inst{1} \and David Schiminovich\inst{2} \and Guinevere Kauffmann\inst{1}
 \and Silvia Fabello\inst{1} \and Cameron Hummels\inst{2} \and Jenna Lemonias\inst{2} 
 \and Sean M. Moran\inst{3} \and Ronin Wu\inst{4} \and Andrew Cooper\inst{1}
 \and Jing Wang\inst{1}
}

\institute{
     Max-Planck Institut f\"{u}r Astrophysik, D-85741 Garching, Germany; \email{bcatinel@mpa-garching.mpg.de}
\and Department of Astronomy, Columbia University, New York, NY 10027, USA
\and Department of Physics and Astronomy, The Johns Hopkins University, Baltimore, MD 21218, USA
\and Commissariat \`a l'Energie Atomique (CEA), 91191 Gif-sur-Yvette, France}

\date{}

\abstract {
We present the second data release from the GALEX Arecibo SDSS
Survey (GASS), an ongoing large Arecibo program to measure the
\hi\ properties for an unbiased sample of \about 1000 galaxies with
stellar masses greater than $10^{10}$ \Msun\ and redshifts
$0.025<z<0.05$. GASS targets are selected from the Sloan Digital Sky
Survey (SDSS) spectroscopic and Galaxy Evolution Explorer (GALEX)
imaging surveys, and are observed until detected or until a gas mass
fraction limit of a few per cent is reached. This second data installment
includes new Arecibo observations of 240 galaxies, and marks the 50\%
of the complete survey. We present catalogs of the \hi, optical and
ultraviolet parameters for these galaxies, and their \hi-line profiles. 
Having more than doubled the size of the sample
since the first data release, we also revisit the main scaling
relations of the \hi\ mass fraction with galaxy stellar mass, stellar mass
surface density, concentration index, and \nuvr\ color, as well as the gas
fraction plane introduced in our earlier work.
}


\titlerunning{The GALEX Arecibo SDSS Survey. VI.}
\authorrunning{Catinella et al.}	

\maketitle

\section{Introduction}\label{s_intro}

Studies of atomic hydrogen (\hi) in galaxies have proved to be of great
importance in order to gain insights into some of the main physical
processes that drive galaxy evolution 
(\eg, reviews by \citealt{roberts94} and \citealt{sancisi08}; 
see also, \eg\ \citealt{things}).
In particular, quantifying how the gas content varies with star
formation and structural properties of galaxies is of paramount
importance for constraining models of galaxy formation.
Equally important is to perform such studies on large
and unbiased samples of galaxies, in order to obtain results
that are truly representative of the local population.
In the past few years we have been carrying out the GALEX Arecibo
SDSS Survey \citep[GASS;][hereafter Paper 1]{gass1}, which is designed 
to provide such a representative sample for massive galaxies, and
whose aim is to understand the role played by gas in the transition
between blue, star-forming galaxies and red, passively-evolving systems. 

GASS uses the Arecibo telescope to measure the \hi\ properties
of \about 1000 galaxies with stellar masses greater than
$10^{10}$ \Msun\ and redshifts $0.025 < z < 0.05$.
For these galaxies, we have homogeneous measurements of structural
parameters from the Sloan Digital Sky Survey \citep[SDSS;][]{sdss},
and ultraviolet (UV) photometry from GALEX \citep{galex} imaging. The availability of
multi-wavelength data is essential in order to connect the atomic gas
to the other galaxy components, and GASS
is optimally configured for follow-up with a range of different
telescopes. At $0.025 < z <0.05$,
the angular diameters of GASS galaxies are small enough that
accurate total CO fluxes can be obtained in a single pointing\footnote{
The FWHM of the IRAM 30m telescope beam is 22\arcsec\ at 115 GHz.
The optical diameters of GASS galaxies, estimated as twice the
Petrosian radius that includes 90\% of the \rband\ light from SDSS,
are all smaller than 1\arcmin, with a mean of 24\arcsec.
}
of the IRAM 30m telescope in the majority of cases \citep[COLD GASS survey,][]{coldgass1}.
Most of the galaxies fit comfortably within a single SDSS frame
and GALEX pointing, so that accurate photometry (and hence stellar masses
and star formation rates) can be measured. The redshift range
does mean, however, that a wide-area blind, shallow survey such as the
Arecibo Legacy Fast ALFA \citep[ALFALFA;][]{alfalfa} survey
only detects the most \hi-rich galaxies. It has thus been necessary to target 
galaxies not detected by ALFALFA in order to measure \hi\ mass fractions
down to a limit of \about 2-5\%. 

The combination of GASS on Arecibo, the COLD GASS follow-up on the IRAM 30m
telescope \citep{coldgass1}, and long-slit optical spectroscopy on the MMT
telescope \citep{moran10,gass5} has yielded a wealth of scientific results. 
We quantified the scaling relations between atomic and molecular
gas fractions and global galaxy properties such as stellar mass,
stellar mass surface density \must, \nuvr\ color and concentration
parameter \citep[Paper~1;][]{coldgass1}.
We showed that galaxies that are unusually \hi-rich for their color
and \must\ have outer disks that are bluer \citep{gass3}, younger
and more metal poor \citep{moran10,gass5}.
We also investigated scaling relations between atomic and molecular
content and star formation rates \citep{gass2,coldgass2}, 
and baryonic mass-velocity-size relations \citep{gass4}. 
Thanks to our multi-wavelength legacy data set, which provides
physical information about the stars and atomic, molecular and ionized
gas in massive systems, we are gaining significant insight into
differences in the evolutionary states of different galaxies, and
setting important constraints for theoretical modeling efforts
\citep[\eg][]{fu10,lagos11,dave11,coldgass3}.

In this paper we present the second data release of GASS, which marks
50\% of the full survey. We use the improved statistics to revisit the
gas fraction scaling relations explored in Paper~1, and discuss 
apparent deviations from linearity that were not evident in the first
data release sample, which included \about 20\% of the full survey.

All the distance-dependent quantities in this work are computed
assuming $\Omega=0.3$, $\Lambda=0.7$ and $H_0 = 70$ \kmsm. 
AB magnitudes are used throughout the paper.

\section{Sample selection, observations and data processing}\label{s_sample}

Survey design, sample selection, Arecibo observations and data
reduction are described in detail in Paper 1,
thus we only provide a summary here, including relevant updates.

GASS measures the global \hi\ properties of \about 1000 galaxies,
selected uniquely by their stellar mass ($10 < {\rm Log} (M_\star/M_\odot) < 11.5$)
and redshift ($0.025 < z < 0.05$). The galaxies are located
within the intersection of the footprints of the SDSS primary
spectroscopic survey, the projected GALEX Medium
Imaging Survey and ALFALFA. We defined a GASS {\it parent sample},
based on SDSS DR6 \citep{sdss6} and the final ALFALFA
footprint, which includes 12006 galaxies that meet our survey
criteria. The targets for 21cm observations are chosen 
by randomly selecting a subset of the parent sample which balances the
distribution across stellar mass and which maximizes existing GALEX
exposure time. 

We observe the galaxies with the Arecibo radio telescope until
we detect them or until we reach a limit of a few percent in
gas mass fraction (defined as \Mhi/\Mst\ in this work). Practically, we have
set a limit of $M_{\rm HI}/M_\star > 0.015$ for galaxies with 
${\rm Log} (M_\star/M_\odot) >10.5$, and a constant gas mass limit 
${\rm Log} (M_{\rm HI}/M_\odot) =8.7$ for galaxies with smaller stellar masses. This
corresponds to a gas fraction limit $0.015-0.05$ for the whole sample.
Given the \hi\ mass limit assigned to each galaxy (set by its gas
fraction limit and stellar mass), we computed the observing time,
\tmax, required to reach that value with our observing mode and
instrumental setup (see below). We exclude from our sample any
galaxies requiring more than 3 hours of total integration time (this
effectively behaves like a redshift cut at the lowest stellar masses).
As mentioned in Paper 1, we do not re-observe galaxies with good
detections already available from ALFALFA and/or the Cornell
\hi\ digital archive \citep[][hereafter S05]{springob05}, a homogeneous
compilation of \hi\ parameters for \about 9000 optically-selected
galaxies.

GASS observations started in March 2008 and are expected to be
completed in 2012. Together with the first data release (DR1, Paper 1),
the data published in this paper amount to \about 50\% of the final
survey sample, and were obtained by the end of February 2011, with a
total allocation of 572 hours of telescope time (of which
\about 13\% unusable due to radio frequency interference [RFI] or
other technical problems).
Arecibo observations are carried out remotely in standard
position-switching mode (\ie\ each observation consists of an on/off
source pair, each typically integrated for 5 minutes, followed
by the firing of a calibration noise diode). We use
the L-band wide receiver and the interim correlator, and 
record the spectra every second with 9-level sampling.
Two correlator boards with 12.5 MHz bandwidth, one polarization, 
and 2048 channels per spectrum (yielding a velocity resolution
of 1.4 \kms\ at 1370 MHz before smoothing) are centered
at or near the frequency corresponding to the SDSS redshift
of the target; two other boards are used for RFI monitoring.

The data reduction, performed in the IDL environment, includes
the following steps (for each on/off pair and polarization): Hanning
smoothing, bandpass subtraction, RFI excision, and flux calibration.
The spectra obtained from each pair are weighted by 1/$rms^2$,
where $rms$ is the root mean square noise measured in the
signal-free portion of the spectrum, and co-added. The two
orthogonal linear polarizations are inspected (if present,
polarization mismatches are noted in Appendix~\ref{s_notes}) and averaged.
The final spectrum is boxcar smoothed, baseline subtracted
(we fitted a low-order polynomial, $n\leq 3$, for 80\% of our sample;
only 6\% of the spectra required $5 \leq n \leq 8$), and
measured as described in Paper 1. The only difference with respect to
DR1 is the estimate of the instrumental broadening correction for the
velocity widths. Measured \hi\ linewidths, \whi, are corrected as follows:
\begin{equation}
        W_{50}^c = \frac{W_{50} -\Delta s}{1+z}
\end{equation}
\noindent
where $z$ is the galaxy redshift and \ds\ is the instrumental
broadening correction, which for DR1 was taken to be the final velocity
resolution $\Delta v$ of the spectrum after smoothing (i.e., between 5 and 21 \kms).
As explained in \citet[][section 2.1]{gass4}, we decided to adopt
$\Delta s = 0.5 \Delta v$, which is in better agreement with other work
\citep[\eg][]{courtois09} and with our own tests on high
signal-to-noise GASS \hi\ profiles. Updated values for the DR1
linewidths can be simply obtained from Table 2 of Paper 1 by applying
equation~1 to \whi\ (column 7) with $\Delta s = 0.5 \Delta v$ (column 5).

\section{The second data release}\label{s_dr2}

This second data release is incremental over DR1, and includes new
Arecibo observations of 240 galaxies. Here we present optical, UV
and 21 cm parameters for these objects, and illustrate the main
characteristics of the combined DR1+DR2 sample in the following section.
The catalogs presented below are available for the combined
DR1 and DR2 samples on the GASS website\footnote{
{\em http://www.mpa-garching.mpg.de/GASS/data.php}
}.

\subsection{SDSS and GALEX data}\label{s_sdss}

This section summarizes the quantities derived from optical and UV
data used in this paper. All the optical parameters listed below were
obtained from Structured Query Language (SQL) queries to the SDSS
DR7 database server\footnote{
{\em http://cas.sdss.org/dr7/en/tools/search/sql.asp}
},
unless otherwise noted.

The GALEX UV photometry for our sample was reprocessed by
us, as explained in \citet{jing10} and summarized in Paper 1.
Briefly, we produced \nuvr\ images by registering GALEX and SDSS
frames, and convolving the latter to the UV point spread
function. The measured \nuvr\ colors are corrected for Galactic
extinction following \citet{wyder07}, from which we obtained
$A_{NUV}-A_r = 1.9807 A_r$ (where the extinction $A_r$ is available
from the SDSS data base and reported in Table~\ref{t_sdss} below). We
do not apply internal dust attenuation corrections.

Table~\ref{t_sdss} lists the relevant SDSS and UV quantities for the GASS
objects published in this work, ordered by increasing right ascension:\\
Cols. (1) and (2): GASS and SDSS identifiers. \\
Col. (3): UGC \citep{ugc}, NGC \citep{ngc} or IC \citep{ic,ic2}
designation, or other name, typically from
the Catalog of Galaxies and Clusters of Galaxies \citep[CGCG;][]{cgcg}, 
or the Virgo Cluster Catalog \citep[VCC;][]{vcc}.\\
Col. (4): SDSS redshift, $z_{\rm SDSS}$. The typical uncertainty of
SDSS redshifts for this sample is 0.0002.\\
Col. (5): base-10 logarithm of the stellar mass, \Mst, in solar
units. Stellar masses are derived from SDSS photometry using the
methodology described in \cite{salim07} (a \citealt{chabrier03}
initial mass function is assumed).
Over our required stellar mass range, these values are
believed to be accurate to better than 30\%.\\
Col. (6): radius containing 50\% of the Petrosian flux in \zband, \Rinz,
in arcsec.\\
Cols. (7) and (8): radii containing 50\% and 90\% of the Petrosian
flux in \rband, $R_{50}$ and  $R_{90}$ respectively, in arcsec (for
brevity, we omit the subscript ``$r$'' from these quantities
throughout the paper).\\
Col. (9): base-10 logarithm of the stellar mass surface density, \must, in
\Msun~kpc$^{-2}$. This quantity is defined as 
$\mu_\star = M_\star/(2 \pi R_{50,z}^2)$, with \Rinz\ in kpc units.\\
Col. (10): Galactic extinction in \rband, ext$_r$, in magnitudes, from SDSS.\\
Col. (11): \rband\ model magnitude from SDSS, $r$, corrected for Galactic extinction.\\
Col. (12): minor-to-major axial ratio from the exponential
fit in \rband, $(b/a)_r$, from SDSS.\\
Col. (13): inclination to the line-of-sight, in degrees, computed as follows:
\begin{equation}
   {\rm cos}~i = \sqrt {\frac{(b/a)^2-q_0^2}{1-q_0^2}} \,,
\end{equation}
\noindent
where $b/a$ is listed in the previous column, and $q_0$ is the
intrinsic axial ratio of a galaxy seen edge-on. We adopt $q_0=0.20$
and set the inclination to 90\deg\ for galaxies with $b/a < 0.2$ (see
\citealt{gass4} and discussion therein). However we provide also
$(b/a)_r$ to allow different estimates of the inclination.\\
Col. (14): \nuvr\ observed color from our reprocessed photometry,
corrected for Galactic extinction.\\
Col. (15): exposure time of GALEX NUV image, T$_{NUV}$, in seconds.\\
Col. (16): maximum on-source integration time, \tmax, required to
reach the limiting \hi\ mass fraction, in minutes (see \S~\ref{s_sample}).
Given the \hi\ mass limit and redshift of each galaxy, \tmax\ is
computed assuming a 5$\sigma$ signal with 300 \kms\ velocity width and
the instrumental parameters typical of our observations (\ie, gain
\about 10 K Jy\minusone\ and system temperature \about 28 K at 1370 MHz).

\subsection{\hi\ source catalogs}\label{s_hi}

The DR2 sample includes 133 detections and 107 non-detections, for
which we provide upper limits below.

The measured \hi\ parameters for the detected
galaxies are listed in Table~\ref{t_det}, ordered by increasing right ascension:\\
Cols. (1) and (2): GASS and SDSS identifiers. \\
Col. (3): SDSS redshift, $z_{\rm SDSS}$. \\
Col. (4): on-source integration time of the Arecibo
observation, $T_{\rm on}$, in minutes. This number refers to
{\it on scans} that were actually combined, and does not account for
possible losses due to RFI excision (usually negligible). \\
Col. (5): velocity resolution of the final, smoothed spectrum in \kms. \\
Col. (6): redshift, $z$, measured from the \hi\ spectrum.
The error on the corresponding heliocentric velocity, $cz$, 
is half the error on the width, tabulated in the following column.\\
Col. (7): observed velocity width of the source line profile
in \kms, \whi, measured at the 50\% level of each peak. 
The error on the width is the sum in quadrature of the 
statistical and systematic uncertainties in \kms. Statistical errors
depend primarily on the signal-to-noise of the \hi\ spectrum, and are
obtained from the rms noise of the linear fits to the edges of the
\hi\ profile. Systematic errors depend on the subjective choice of the
\hi\ signal boundaries (see Paper 1), and are negligible for most of
the galaxies in our sample (see also Appendix~\ref{s_notes}).\\
Col. (8): velocity width corrected for instrumental broadening
and cosmological redshift only, \whi$^c$, in \kms\ (see equation
1). No inclination or turbulent motion corrections are applied.\\
Col. (9): observed, integrated \hi-line flux density in Jy \kms,
$F \equiv \int S~dv$, measured on the smoothed and baseline-subtracted
spectrum. The reported uncertainty is the sum in quadrature of the 
statistical and systematic errors (see col. 7).
The statistical errors are calculated according to equation 2 of S05:
\begin{displaymath}
 \epsilon^{stat} = 2 ~rms \sqrt{1.4 W_{50} \Delta v},
\end{displaymath}
\noindent
where $rms$ is the noise measured in the signal-free part of the
spectral baseline (see col. 10), $\Delta v$ is the velocity resolution 
of the smoothed spectrum (see col. 5), and the factor 2 accounts for
the contribution from uncertainties in the baseline fit (following 
\citealt{schneider90}).\\
Col. (10): rms noise of the observation in mJy, measured on the
signal- and RFI-free portion of the smoothed spectrum.\\
Col. (11): signal-to-noise ratio of the \hi\ spectrum, S/N,
estimated following \citet{saintonge07} and adapted to the velocity
resolution of the spectrum. 
This is the definition of S/N adopted by ALFALFA, which accounts for the
fact that for the same peak flux a broader spectrum has more signal.\\
Col. (12): base-10 logarithm of the \hi\ mass, \Mhi, in solar
units, computed via: 
\begin{equation}
    \frac{M_{\rm HI}}{\rm M_{\odot}} = \frac{2.356\times 10^5}{1+z}
    \left[ \frac{d_{\rm L}(z)}{\rm Mpc}\right]^2
    \left(\frac{\int S~dv}{\rm Jy~km~s^{-1}} \right)
\label{eq_MHI}
\end{equation}
\noindent
where $d_{\rm L}(z)$ is the luminosity distance to the galaxy at
redshift $z$ as measured from the \hi\ spectrum. \\
Col. (13): base-10 logarithm of the \hi\ mass fraction, \Mhi/\Mst.\\
Col. (14): quality flag, Q (1=good, 2=marginal, 3=marginal and
confused, 5=confused). An asterisk indicates the presence of a note
for the source in Appendix~\ref{s_notes}.
Code 1 indicates reliable detections, with a S/N ratio of order of
6.5 or higher (this is the same threshold adopted by ALFALFA). 
Marginal detections have lower S/N, thus more uncertain
\hi\ parameters, but are still secure detections, with \hi\ redshift
consistent with the SDSS one.
The S/N limit is not strict, but depends also on \hi\ profile and baseline
quality. As a result, galaxies with S/N slightly above the threshold
but with uncertain profile or bad baseline may be flagged with a
code 2, and objects with S/N $\lesssim 6.5$ and \hi\ profile with
well-defined edges may be classified as code 1. 
We assigned the quality flag 5 to eighteen ``confused'' galaxies, where
most of the \hi\ emission is believed to come from another source
within the Arecibo beam. For some of the galaxies, the presence of
small companions within the beam might contaminate (but is unlikely to
dominate) the \hi\ signal -- this is just noted in Appendix~\ref{s_notes}.
Finally, we assigned code 3 to twelve galaxies, which are both marginal and
confused.\\

Table~\ref{t_ndet} gives the derived \hi\ upper limits for the non-detections. 
Columns (1-4) and (5) are the same as columns (1-4) and (10) in Table~\ref{t_det},
respectively. Column (6) lists the upper limit on the \hi\ mass in
solar units, Log \Mhi$_{,lim}$, computed assuming a 5 $\sigma$ signal with 300 \kms\ 
velocity width, if the spectrum was smoothed to 150 \kms. Column (7)
gives the corresponding upper limit on the gas fraction, Log~\Mhi$_{,lim}$/\Mst.   
An asterisk in Column (8) indicates the presence of a note for the
galaxy in Appendix~\ref{s_notes}.\\

SDSS images and \hi\ spectra of the DR2 galaxies are presented in
Appendix~\ref{s_spectra}, organized as follows: \hi\ detections with quality
flag 1 in Table~\ref{t_det} (Figure~\ref{det}), marginal detections
with quality flag 2 (Figure~\ref{dcode2}) and 3 (Figure~\ref{dcode3}),
confused detections (Figure~\ref{dcode5})
and non-detections (Figure~\ref{ndet}).
The objects in each of these figures are ordered by 
increasing GASS number (indicated on the top right corner of each spectrum).
The SDSS images show a 1 arcmin square field, \ie\ only the central
part of the region sampled by the Arecibo beam (the half
power full width of the beam is \about 3.5\arcmin\ at the
frequencies of our observations). Therefore, companions that might be
detected in our spectra typically are not visible in the
postage stamps, but they are noted in Appendix~\ref{s_notes}.
The \hi\ spectra are always displayed over a 3000 \kms\ velocity
interval, which includes the full 12.5 MHz bandwidth adopted for our
observations. The \hi-line profiles are calibrated, smoothed 
(to a velocity resolution between 5 and 21 \kms\ for
the detections, as listed in Table~\ref{t_det}, or to
\about 15 \kms\ for the non-detections), and
baseline-subtracted. A red, dotted line indicates the heliocentric
velocity corresponding to the optical redshift from SDSS. 
In Figures~\ref{det}-\ref{dcode5}, the shaded area and two vertical
dashes show the part of the profile that was integrated to
measure the \hi\ flux and the peaks used for width measurement,
respectively.

\begin{figure}
\begin{center}
\includegraphics[width=9cm]{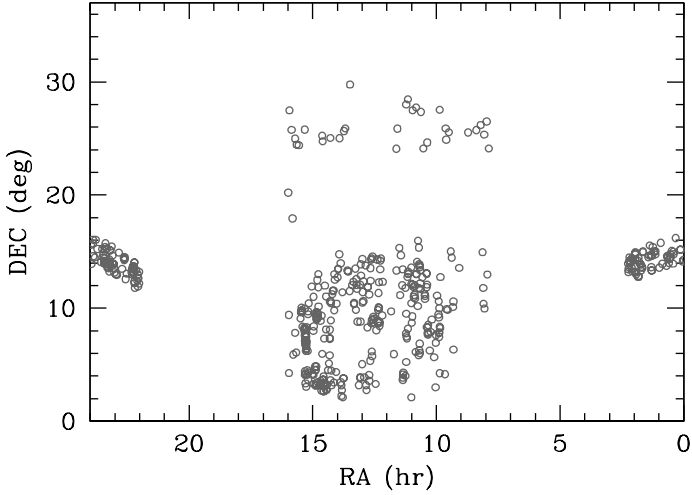}
\caption{Sky distribution of the GASS representative sample (480 galaxies).}
\label{skyd}
\end{center}
\end{figure}

\section{GASS sample properties}\label{s_properties}

Taken together, the first and second GASS data releases include
416 galaxies, of which 232 are \hi\ detections and 184 are
non-detections. We will refer to this as the GASS {\it observed}
sample. Because we do not reobserve galaxies with good \hi\ detections
already available from either ALFALFA or the S05 archive, this data
set lacks the most gas-rich objects, which need to be added back 
in the correct proportions. By following the procedure
described in section 7.2 of Paper 1, we obtained a sample that
includes 480 galaxies (of which 296 are detections) and that is
representative in terms of \hi\ properties. We will
refer to this as the GASS {\it representative} sample. Notice that,
because of the improved statistics compared to DR1, here
we use only one such representative sample (as opposed to a
suite of 100 realizations, differing for the set of randomly-selected
gas-rich galaxies added to the GASS observations).

The sky distribution of the representative sample is shown in
Figure~\ref{skyd}. We restricted the observations as much as possible
to the two declination intervals from $+4$\deg\ to $+16$\deg\ and 
from $+24$\deg\ to $+28$\deg, for which ALFALFA catalogs were
available to us in advance of publication
\citep{alfalfa40}. The uneven right ascension distribution is the
result of telescope allocation, which favored small observing session
at LST intervals less oversubscribed (such as 14$-$16 hours). 

\begin{figure*}
\begin{center}
\includegraphics[width=18cm]{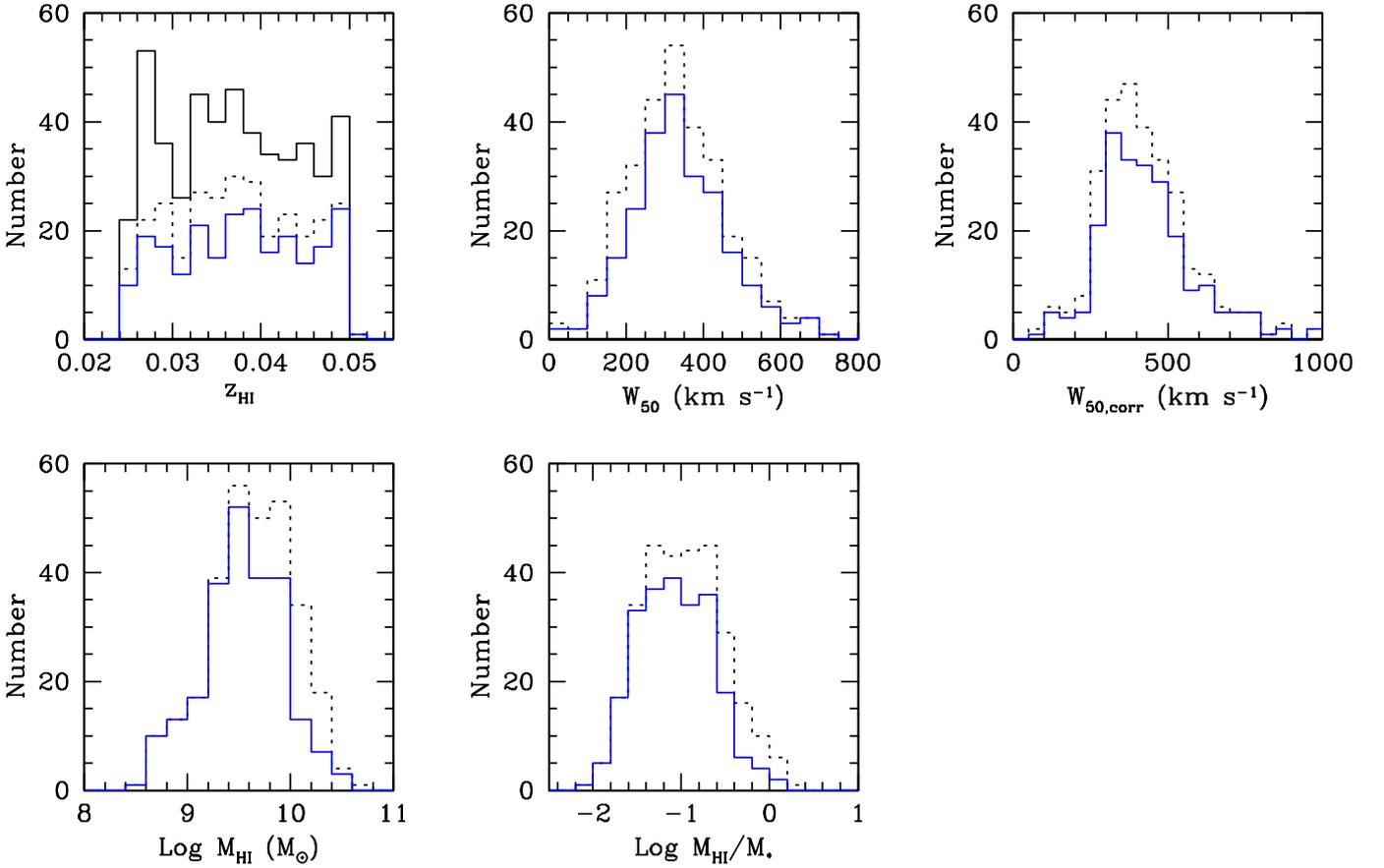}
\caption{Distributions of redshifts, velocity widths, 
velocity widths corrected for inclination,
\hi\ masses and gas mass fractions for the galaxies with
\hi\ detections from GASS (blue histograms, 232 galaxies).
Dotted histograms correspond to the representative sample, which
includes gas-rich objects from ALFALFA and/or S05 archive (see text).
The solid black histogram in the top left panel shows the
distribution of SDSS redshifts for the full 
sample (\ie\ including the non-detections).}
\label{hi}
\end{center}
\end{figure*}

The \hi\ properties of the detected galaxies are illustrated in
Figure~\ref{hi} for both observed (blue histograms) and
representative (dotted) samples. The solid black histogram in the top
left panel shows the redshift distribution for the full representative
sample, using the SDSS redshifts for the non-detections. As for the
DR1 sample presented in Paper 1, the distribution of corrected
velocity widths (which have not been deprojected to edge-on view)
peaks near 300 \kms, which is the value that we assume to compute
upper limits for the \hi\ masses of the non-detections, and to
estimate \tmax\ in Table~\ref{t_sdss}. 

\begin{figure*}
\begin{center}
\includegraphics[width=18cm]{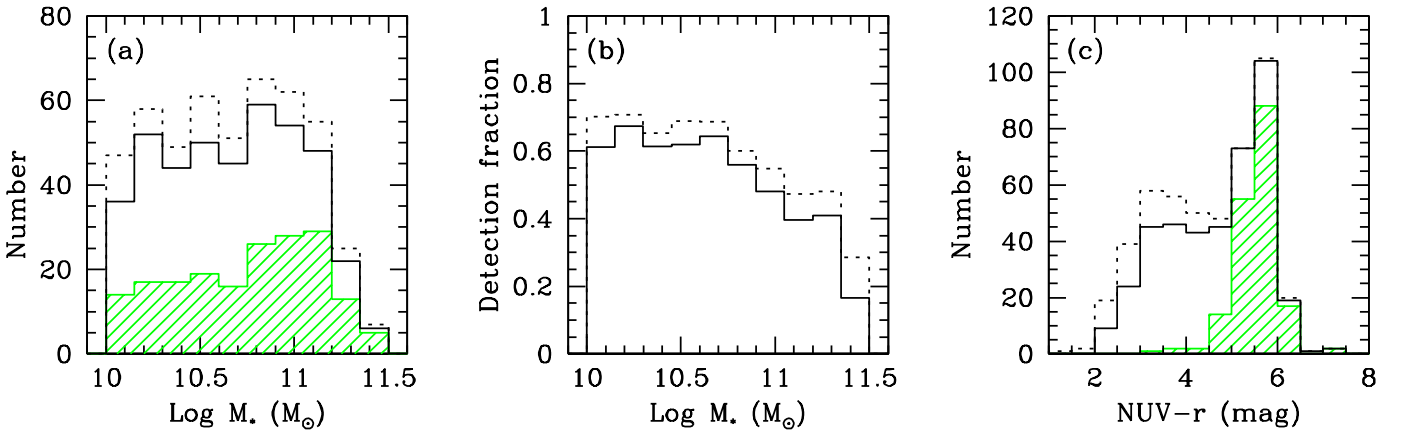}
\caption{Stellar mass (a) and observed \nuvr\ color (c)
distributions for the GASS observed sample (solid). Hatched
histograms indicate the corresponding distributions for the
non-detections. The detection fraction (\ie, the ratio of detections to total) is
shown as a function of stellar mass in (b). The dotted histograms in all
panels show the distributions for the representative sample (see text).}
\label{opt}
\end{center}
\end{figure*}

Figure~\ref{opt} presents the stellar mass (a) and \nuvr\ color (c)
distributions for the observed (black histogram) and representative
(dotted) samples. The corresponding distributions for the
non-detections are shown as hatched green histograms. The stellar
mass histogram is almost flat by survey design, as we wish to obtain
similar statistics in each bin in order to perform comparisons at fixed
stellar mass. As already noted
in Paper 1, non-detections span the entire range of stellar masses,
but they are concentrated in the red portion of the \nuvr\ space.
The detection fraction, \ie\ the ratio of detected galaxies to total,
is plotted as a function of stellar mass in (b). The detection
fraction is close to 70\% for \Mst $< 10^{10.7}$ \Msun, and drops to 
\about 30\% in the highest stellar mass bin.

The \nuvr\ color-stellar mass diagram shown in Figure~\ref{cmd}
combines the information contained in Figures~\ref{opt}a and
\ref{opt}c. In order to clearly indicate the loci of the blue cloud
and red sequence, we use the GASS {\it parent sample} mentioned in
\S~\ref{s_sample}, which is the complete set of 12,006 galaxies
that meet our selection criteria. The locations of these galaxies in
the diagram are shown by the grayscales; the red sequence peaks at 
\nuvr \about 5.5 mag, and the blue cloud is mostly confined to 
\nuvr $<3.5$ mag. The results for the GASS representative sample are
indicated by red and green symbols. GASS non-detections are almost
entirely confined to the red sequence. 
We note that, because we select targets with an
approximately flat stellar mass distribution (see Section 2 and
Fig.~\ref{opt}), we oversample the high stellar mass galaxies, which
are more rare in a volume-limited survey. Hence the distribution of our
representative sample in the color-stellar mass diagram is somewhat offset
toward higher \Mst\ values compared to that of the GASS {\it parent sample}.

\section{Gas fraction scaling relations}

The updated versions of the scaling relations
investigated in Paper 1 are presented in Figures~\ref{dr2gf}
and \ref{scalings}. We describe the figures first, and discuss them
together afterward.

Clockwise from the top left, Figure~\ref{dr2gf} shows how the gas mass
fraction \Mhi/\Mst\ depends on stellar mass, stellar mass surface
density, observed \nuvr\ color and concentration index for the GASS
representative sample. Red circles and green upside-down triangles
indicate \hi\ detections and non-detections (plotted at their upper
limits), respectively. ALFALFA detections of galaxies in the parent
sample (1102 objects in total), whose \hi\ masses have been computed
consistently with GASS ones from the fluxes tabulated by
\citet{alfalfa40}, are shown as gray dots for comparison.
Dotted lines in each panel are linear fits to the detections, which we
use only to quantify the scatter $\sigma$.

The average values of the gas fraction are plotted as a function of
the same quantities in Figure~\ref{scalings}. Gray and green symbols
reproduce individual GASS detections and non-detections, respectively,
from Figure~\ref{dr2gf}. The averages are computed including the
non-detections, whose \hi\ masses were set either to their upper
limits (green circles) or to zero (red circles). As in Paper 1, these
averages are weighted in order to compensate for the flat stellar mass
distribution of the GASS sample, using the volume-limited parent
sample as a reference. Briefly, we binned both the parent sample
and the GASS representative sample by stellar mass (with a 0.2 dex
step), and used the ratio between the two histograms as a weight.
Error bars indicate the standard deviation of the weighted averages.
Lastly, weighted median values of the gas fraction, calculated using
upper limits for the \hi\ masses of the non-detections, are plotted as
green triangles. The values of weighted average and median gas
fractions shown in this figure are listed in Table~\ref{t_avgs} for
reference.

These results are consistent with our previous findings
(see also \citealt{fabello1} and \citealt{luca11}), and show that:\\
-- The gas fraction of GASS detections is a decreasing function of
stellar mass, stellar mass surface density, and \nuvr\ color.
The scatters around the linear fits to these relations are unchanged
with respect to DR1 ($\sigma =0.39, 0.36$ and 0.33 dex, respectively),
despite the fact that the sample size has more than doubled (from
\about 200 to 480 galaxies).\\
-- The strongest correlation is with observed \nuvr\ color
(Pearson correlation coefficient $r=-0.69$); the
average \Mhi/\Mst\ decreases from 73\% to 2\% from the bluest to the
reddest galaxies. The decrease quoted in Paper 1 was smaller because,
due to poorer statistics, we did not have enough observations below
\nuvr $=2.9$ mag to compute an average gas fraction.
Given the link between star formation rate and gas content, the strong
correlation between \nuvr\ and gas fraction is expected, though as we
note further below and in the next section, departures from a tight,
linear correlation may result from contributions to UV light from an
older population, or alternatively, diminished UV from dust
attenuation.\\
-- The weakest correlations are with stellar mass ($r=-0.53$)
and concentration index ($r=-0.37$; the scatter is $\sigma
=0.43$ dex, it was 0.45 dex for the DR1 sample). In both cases, the
difference between average and median gas fractions points to the
presence of significant tails of galaxies with small values of
\Mhi/\Mst. The relation between gas fraction and stellar mass
surface density has a correlation coefficient $r=-0.62$.\\
-- All the non-detections have stellar mass surface density
\must $> 10^{8.5}$ \Msun~kpc$^{-2}$. The average gas fractions are
insensitive to the way we treat the non-detections, except for the
very most massive, dense and red galaxies.\\
-- In the GASS stellar mass and redshift intervals, ALFALFA only
detects the bluest, most gas-rich objects.\\

\begin{figure}
\begin{center}
\includegraphics[width=9cm]{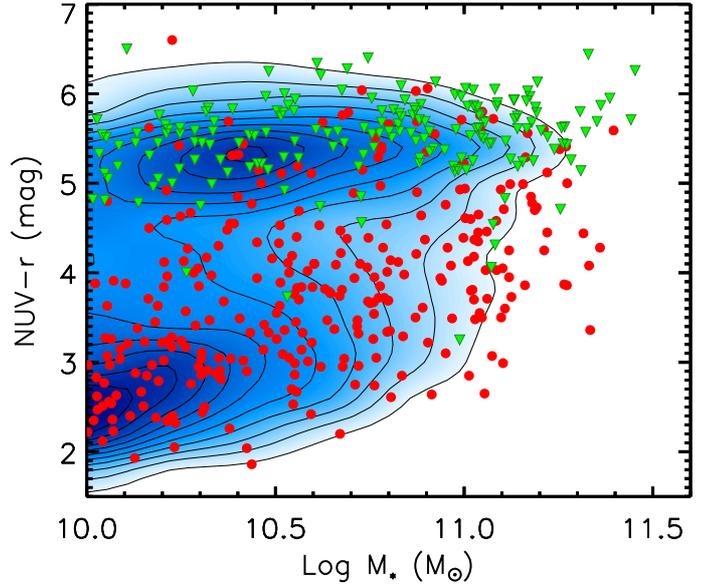}
\caption{Color-stellar mass diagram for the GASS {\it parent sample},
the super-set of \about 12,000 galaxies that meet the survey criteria
(grayscales). Red circles and green upside-down triangles indicate
\hi\ detections and non-detections, respectively, from the
representative sample.}
\label{cmd}
\end{center}
\end{figure}

\begin{figure*}
\begin{center}
\includegraphics[width=18cm]{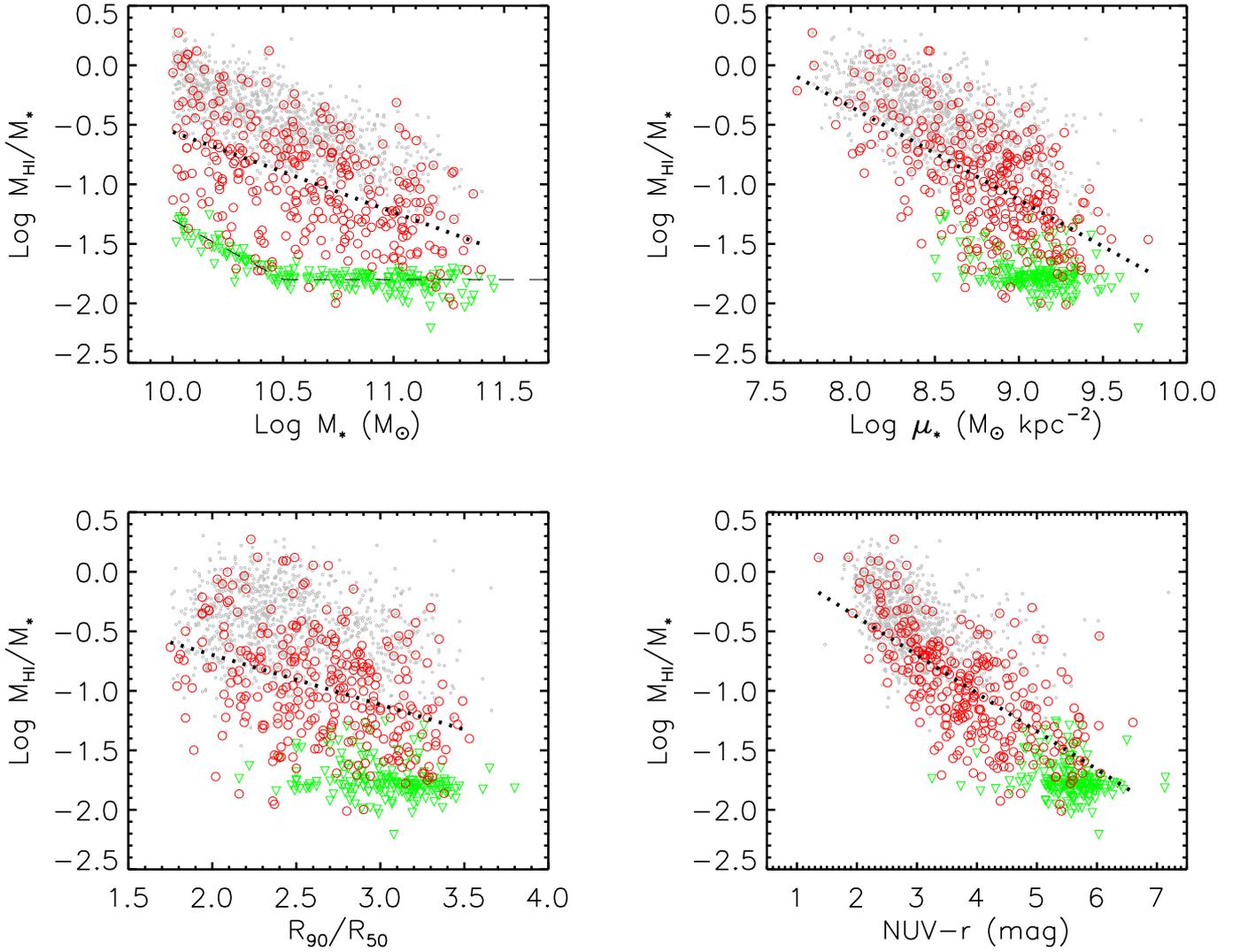}
\caption{The \hi\ mass fraction of the GASS sample is plotted here as
a function of stellar mass, stellar mass surface density,
concentration index, and observed \nuvr\ color. Red circles and green triangles
represent detections and non-detections, respectively. 
For comparison, we also show the full set of ALFALFA galaxies meeting the GASS
selection criteria that have been cataloged to date (gray). 
The dashed line on the top-left panel indicates the \hi\ detection limit
of the GASS survey. Dotted lines in each panel are linear fits to
the \hi\ detections only.}
\label{dr2gf}
\end{center}
\end{figure*}

\begin{figure*}
\begin{center}
\includegraphics[width=18cm]{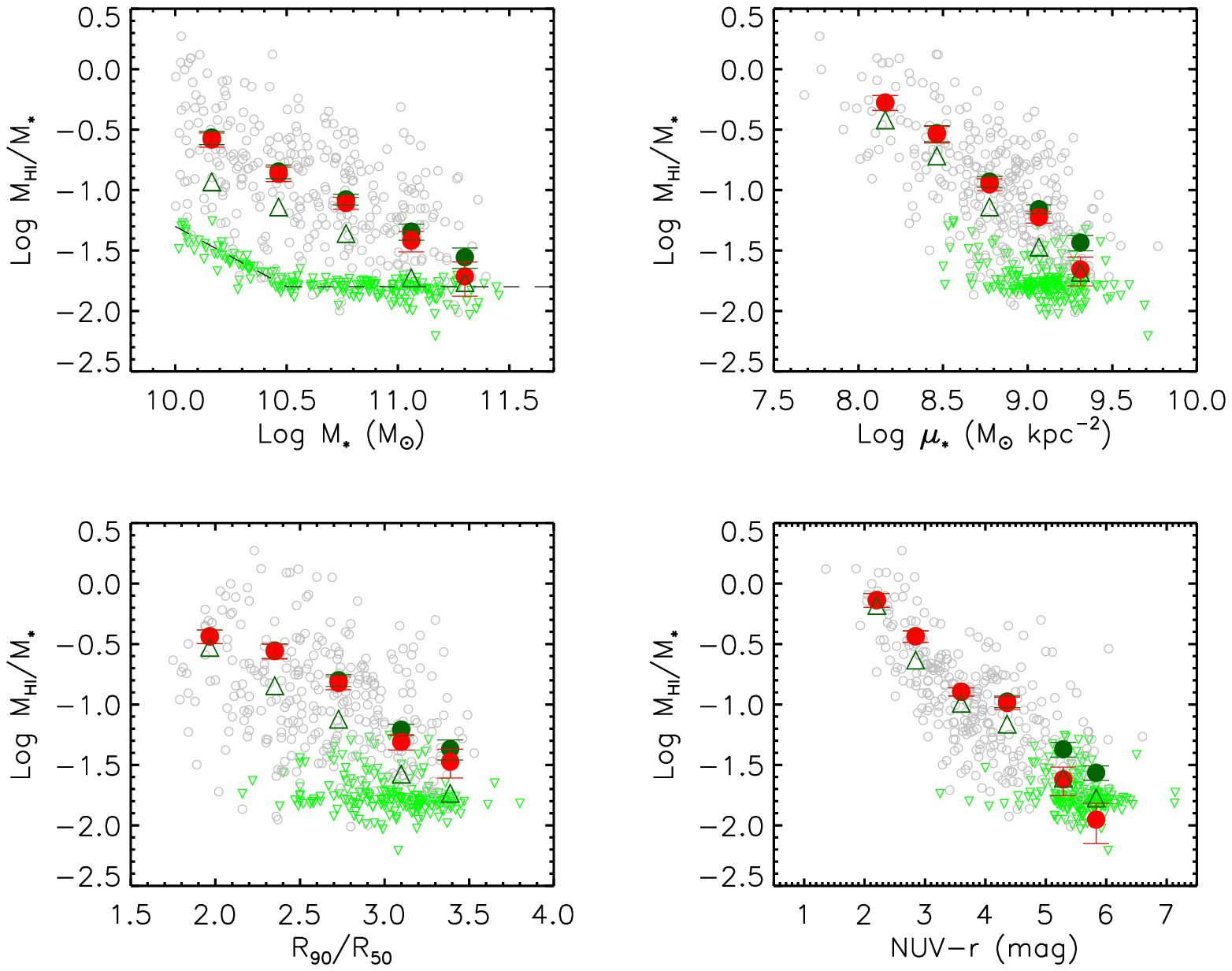}
\caption{Average trends of \hi\ mass fraction as a function of stellar
mass, stellar mass surface density, concentration index and observed
\nuvr\ color for the representative sample. 
In each panel, large circles indicate weighted average gas fractions 
(see text). These were computed including the
non-detections, whose \hi\ mass was set to either its upper limit
(dark green) or to zero (red). Green triangles are weighted medians.
Only averages based on at least 8 galaxies are shown. These results
are listed in Table~\ref{t_avgs}. 
GASS data from Figure~\ref{dr2gf} are shown in gray and green (for detections
and non-detections, respectively). The dashed line in the first panel
shows the \hi\ detection limit of the GASS survey.}
\label{scalings}
\end{center}
\end{figure*}

With our improved statistics, the relation between gas fraction and
\nuvr\ color {\it for the detections} now seems to show a break at 
\nuvr \about 3.5 mag, which corresponds approximately to the upper
envelope of the blue sequence (see Figure~\ref{cmd}). The galaxies
with the highest gas fractions systematically lie above the linear
fit to the detections (Figure~\ref{dr2gf}). This can be seen also in 
Figure~\ref{scalings}, if we focus on the bins that are not
dominated by the non-detections. Interestingly, including the
non-detections in the computation of the average gas fractions
restores the linearity of the relation with \nuvr\ color. This
agrees with the results based on the Herschel Reference Survey 
\citep[HRS;][]{hrs} sample, which is more local and thus
probes significantly lower gas fractions than GASS.
The HRS \Mhi/\Mst\ versus \nuvr\ relation does not show any
evidence for a change of slope \citep[see Fig. 1 of][]{luca11}, 
but only an increase of scatter outside the blue sequence,
which we observe as well.
We will come back to this point in the next section.

\section{Gas fraction plane}

\begin{figure*}
\begin{center}
\includegraphics[width=17.5cm]{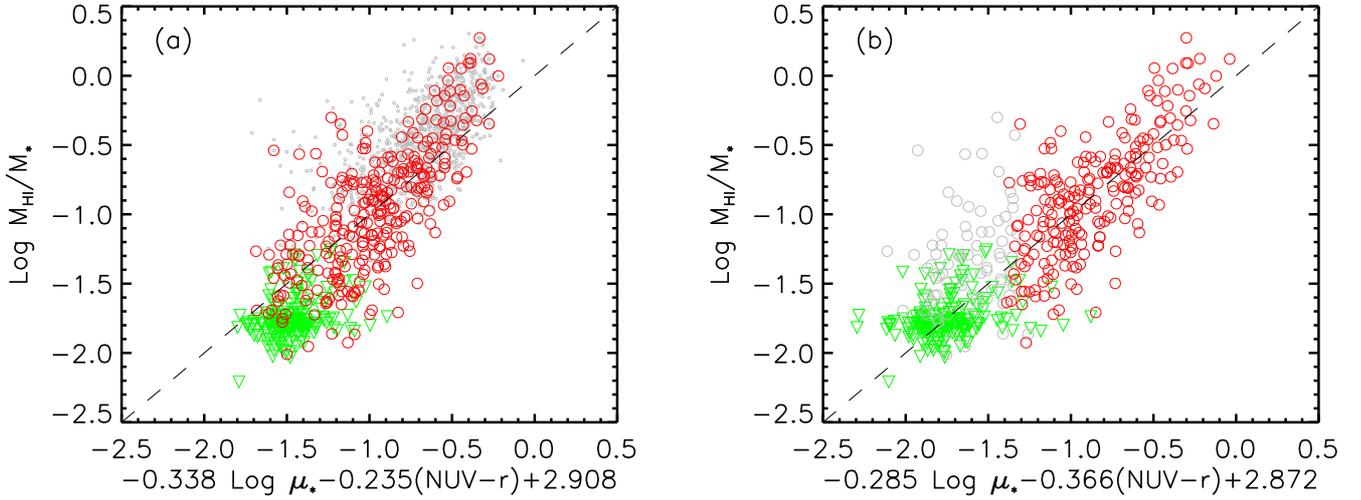}
\caption{
Gas fraction plane, a relation between \hi\ mass fraction and a
linear combination of stellar mass surface density and observed
\nuvr\ color. (a) Relation obtained using all the \hi\ detections in the
GASS representative sample (red circles). The symbols are the same as those in Fig.~\ref{dr2gf}.
(b) Relation obtained using only the subset of detected galaxies with
\nuvr $\leq 4.5$ mag (red circles). Gray circles and green upside-down
triangles indicate the remaining \hi\ detections and the
non-detections, respectively.}
\label{plane}
\end{center}
\end{figure*}

In Paper 1 we introduced the {\it gas fraction plane}, a relation
between gas mass fraction and a linear combination of \nuvr\ color
(which is a proxy for star formation rate per unit stellar mass) and
stellar mass surface density, which can be used to define what is
``\hi\ normalcy'' for local massive, star-forming galaxies. As
discussed by \cite{zhang09}, such a relation is a direct consequence
of the Kennicutt-Schmidt global star formation law 
\citep{schmidt63,kennicutt98} if one assumes that star formation and gas
densities are computed over the same spatial area --- thus it is
physically motivated. 
Figure~\ref{plane}a shows the result for the GASS
representative sample. We remind the reader that the best fit relation
is obtained following \citet{bernardi03} (\ie, this is the solution
that minimizes the scatter on the $y$ coordinate, and therefore it is
equivalent to a direct fit), and that only \hi\ detections
(red circles) are used. The coefficients of the fit (reported on the
$x$ axis of the figure) and the rms scatter in Log \Mhi/\Mst, 0.319 dex,
are almost unchanged with respect to the DR1 solution, showing that
the 20\% survey sample was indeed representative. The Pearson
correlation coefficient for the relation shown in Figure~\ref{plane}a is $r=-0.71$.

We note that \citet{cheng12} presented a slightly different solution
for this gas fraction plane (see their Fig. 2), obtained by weighting
each galaxy by the mass-dependent selection function of GASS. However,
the weights make negligible difference to the result (they obtain 
Log \Mhi/\Mst $=-0.322$ Log \must $-0.234$ \nuvr $+2.817$, with
identical scatter, 0.32 dex).

Our gas fraction plane is also consistent with the one based on the
HRS sample, when this is restricted to the \hi -normal galaxies
(\ie, when \hi -deficient systems in the Virgo cluster are
excluded; see \citealt{luca11}). This agreement is a
non-trivial result, given the different selection
criteria and data sets of the two surveys. Notice however that the
scatter of the HRS plane is smaller (0.27 dex, see below).

As can be seen in Figure~\ref{plane}a, the highest gas fraction
galaxies lie systematically above the mean relation, indicated by a
dashed line. This is the same deviation from linearity discussed for
the \Mhi/\Mst\ versus \nuvr\ relation in the previous section. 
As already noted, the apparent break of the gas fraction-color
relation seen for the \hi\ detections is linked to the gas fraction
limit of GASS. Indeed the break effectively disappears when 
non-detections are included in the averages, and a similar
discontinuity is not seen for the HRS sample, which probes 
significantly lower gas fractions.

Because the gas fraction plane is computed using only
detections, its validity breaks down in the region where the
contribution of the non-detections becomes important. Thus, a more
reliable solution for the plane can be obtained by
using only galaxies with \nuvr $\leq 4.5$ mag, where we have
virtually only detections (this is a conservative threshold based on
the inspection of Figure~\ref{opt}c). This cut has
the additional advantage of excluding a region of parameter space
that is problematic for two reasons.
First, outside the blue sequence, the UV emission might 
not be physically associated to the \hi, but might trace a
more evolved stellar population, and thus \nuvr\ might no longer be a good proxy
for specific star formation rate (see also \citealt{oconnell99,boselli05,lucatom09}).
This increases the scatter of the gas fraction-color relation outside
the blue sequence.
Second, both \nuvr\ and \must\ saturate, \ie\ they never exceed
\nuvr \about 6 mag and Log \must \about 9.5 \Msun\ kpc$^{-2}$. This 
will introduce an apparent non-linearity in the relations involving
gas fraction regardless of \hi\ content.

The gas fraction plane computed using only galaxies with 
\nuvr $\leq 4.5$ mag is shown in Figure~\ref{plane}b: the relation
becomes more linear, but clearly the scatter of the redder galaxies
(shown in gray) increases. Importantly, the main outliers 
remain the same. The scatter of the plane in Figure~\ref{plane}b
is 0.29 dex over the subset of galaxies used for its computation
(and 0.32 dex when all detections are included), which better agrees
with the scatter of the HRS plane (0.27 dex).

We note that the GASS DR1 sample included too few galaxies on the top right corner
of the plane to notice a clear deviation for the galaxies with the highest gas
fractions. Our improved statistics allows us to look now into second order corrections,
such as the one suggested above. Naturally, one has to keep in mind that
the relation used to predict gas fractions has only a statistical validity, and 
should not be trusted for an individual galaxy.

Another approach to obtain a better prediction for the
gas fractions of the galaxies with the highest values of \Mhi\ and/or
\Mhi/\Mst\ is to correct the non-linearity by adding
degrees of freedom when fitting the plane
(although this is no longer physically justified by the Kennicutt-Schmidt 
star formation law). As shown by \citet{gass3}, gas-rich galaxies
tend to have bluer-than-average outer disks. Thus, \citet{cheng12} advocate that
a new gas fraction estimator that includes two additional parameters,
stellar mass and $g-i$ color gradient (defined as the difference
between outer and inner $g-i$ color), yields a better fit to the
\hi-rich galaxies. However, over the GASS stellar mass regime, the
scatter decreases very little (from 0.32 to 0.31 dex, see their Figure 2).

\begin{table*}
\centering
\caption{Weighted Average and Median Gas Fractions}
\label{t_avgs}
\begin{tabular}{lrcccr}
\hline\hline
\vspace{-6pt}\\
 &  & $\langle M_{\rm HI}/M_\star \rangle$  & $\langle M_{\rm HI}/M_\star \rangle$   & \Mhi/\Mst &  \\
$x$  & $\langle x \rangle$ & (average)$^{a}$ & (average)$^{b}$ & (median)$^{c}$ & $N^{d}$ \\
\vspace{-6pt}\\
\hline
\vspace{-6pt}\\
Log \Mst  &    10.16 &  0.272$\pm$0.034 &  0.262$\pm$0.035 &  0.117 & 105   \\
   	  &    10.46 &  0.143$\pm$0.018 &  0.136$\pm$0.019 &  0.072 & 110   \\
   	  &    10.77 &  0.084$\pm$0.009 &  0.078$\pm$0.009 &  0.044 & 116   \\
   	  &    11.06 &  0.045$\pm$0.007 &  0.038$\pm$0.007 &  0.019 & 117   \\
   	  &    11.30 &  0.028$\pm$0.005 &  0.019$\pm$0.006 &  0.017 &  32   \\
          &          &                  &                  &        &       \\
Log \must &     8.16 &  0.530$\pm$0.075 &  0.530$\pm$0.075 &  0.380 &  24   \\
   	  &     8.46 &  0.296$\pm$0.045 &  0.291$\pm$0.046 &  0.191 &  50   \\
   	  &     8.77 &  0.118$\pm$0.012 &  0.112$\pm$0.012 &  0.072 & 119   \\
   	  &     9.07 &  0.069$\pm$0.006 &  0.060$\pm$0.006 &  0.034 & 193   \\
   	  &     9.31 &  0.037$\pm$0.005 &  0.022$\pm$0.006 &  0.021 &  81   \\
          &          &                  &                  &        &       \\
\cindx	  & 	1.97 &  0.366$\pm$0.047 &  0.366$\pm$0.047 &  0.296 &  33   \\
   	  & 	2.35 &  0.279$\pm$0.039 &  0.276$\pm$0.039 &  0.143 &  91   \\
   	  & 	2.73 &  0.159$\pm$0.018 &  0.150$\pm$0.018 &  0.076 & 137   \\
   	  & 	3.10 &  0.062$\pm$0.007 &  0.049$\pm$0.007 &  0.026 & 176   \\
   	  & 	3.39 &  0.043$\pm$0.008 &  0.034$\pm$0.009 &  0.018 &  42   \\
          &          &                  &                  &        &       \\
\nuvr 	  & 	2.20 &  0.730$\pm$0.096 &  0.730$\pm$0.096 &  0.659 &  15   \\
   	  & 	2.85 &  0.366$\pm$0.040 &  0.366$\pm$0.040 &  0.233 &  65   \\
   	  & 	3.60 &  0.128$\pm$0.010 &  0.127$\pm$0.010 &  0.102 &  91   \\
   	  & 	4.36 &  0.106$\pm$0.012 &  0.103$\pm$0.012 &  0.069 &  79   \\
   	  & 	5.29 &  0.043$\pm$0.006 &  0.024$\pm$0.006 &  0.025 & 123   \\
   	  & 	5.83 &  0.027$\pm$0.004 &  0.011$\pm$0.004 &  0.017 &  94   \\
\vspace{-6pt}\\
\hline\hline
\end{tabular}
\begin{flushleft}
Notes. --- $^{a}$Weighted, average gas fraction; \hi\ mass of non-detections set to upper limit.\\
$^{b}$Weighted, average gas fraction; \hi\ mass of non-detections set to zero.\\
$^{c}$Weighted, median gas fraction; \hi\ mass of non-detections set to upper limit.\\
$^{d}$Number of galaxies in the bin.
\end{flushleft}
\end{table*}


\section{Conclusions}

This paper presents the second data release of GASS, an ongoing large
Arecibo program to measure \hi\ parameters for \about 1000 massive
galaxies, selected from the SDSS spectroscopic and GALEX imaging
surveys. This release is incremental over the first one (Paper~1), and
includes new \hi\ observations for 240 galaxies. The representative
sample presented here, which was obtained by adding the correct
proportion of \hi -rich objects detected by ALFALFA or in the S05
archive that we did not re-observe with Arecibo, includes 480
galaxies, and marks the 50\% of the full survey.

We discussed the properties of the 50\% survey sample, and used it to 
revisit the scaling relations between gas mass fraction and galaxy
structural parameters and color, as well as the gas fraction plane,
presented in Paper~1. Overall our results confirm our previous
findings, which were based on the initial 20\% survey sample, and with
almost identical scatters.
However, the significantly improved statistics also allowed us to
notice second-order effects, in the form of a systematic deviation of
the high gas fraction tail in the \Mhi/\Mst\ versus \nuvr\ and gas
fraction plane relations. We identify the cause for such a deviation
in the sensitivity limit of GASS, which is a gas fraction
limited survey. Above a \nuvr\ color of 4.5 mag, the results are
dominated by the non-detections, which seem to cause an apparent break
in the gas fraction versus color relation. As a result, the gas
fraction plane slightly underpredicts \Mhi/\Mst\ at the high end.
Possible solutions include fitting the plane where the relation
between gas fraction and \nuvr\ is not dominated by the
non-detections, or include additional parameters to the fit.

The new catalogs of \hi, optical and UV parameters presented in this
work increase the legacy value of GASS, and place our investigations
of what physical processes are responsible for the transition between
blue, star-forming galaxies and red, passively-evolving systems on a
statistically more solid ground.\\

\begin{acknowledgements}

We thank the anonymous referee for useful suggestions.

This research has made use of the NASA/IPAC Extragalactic Database
(NED) which is operated by the Jet Propulsion Laboratory, California
Institute of Technology, under contract with the National Aeronautics
and Space Administration.

The Arecibo Observatory is operated by SRI International under a
cooperative agreement with the National Science Foundation
(AST-1100968), and in alliance with Ana G. M{\'e}ndez-Universidad
Metropolitana, and the Universities Space Research Association.

GALEX (Galaxy Evolution Explorer) is a NASA Small Explorer, launched
in April 2003. We gratefully acknowledge NASA's support for
construction, operation, and science analysis for the GALEX mission,
developed in cooperation with the Centre National d'Etudes Spatiales
(CNES) of France and the Korean Ministry of Science and Technology. 

Funding for the SDSS and SDSS-II has been provided by the Alfred
P. Sloan Foundation, the Participating Institutions, the National
Science Foundation, the U.S. Department of Energy, the National
Aeronautics and Space Administration, the Japanese Monbukagakusho, the
Max Planck Society, and the Higher Education Funding Council for
England. The SDSS Web Site is http://www.sdss.org/.

The SDSS is managed by the Astrophysical Research Consortium for the
Participating Institutions. The Participating Institutions are the
American Museum of Natural History, Astrophysical Institute Potsdam,
University of Basel, University of Cambridge, Case Western Reserve
University, University of Chicago, Drexel University, Fermilab, the
Institute for Advanced Study, the Japan Participation Group, Johns
Hopkins University, the Joint Institute for Nuclear Astrophysics, the
Kavli Institute for Particle Astrophysics and Cosmology, the Korean
Scientist Group, the Chinese Academy of Sciences (LAMOST), Los Alamos
National Laboratory, the Max-Planck-Institute for Astronomy (MPIA),
the Max-Planck-Institute for Astrophysics (MPA), New Mexico State
University, Ohio State University, University of Pittsburgh,
University of Portsmouth, Princeton University, the United States
Naval Observatory, and the University of Washington.

\end{acknowledgements}

\bibliography{biblio}


\appendix

\section{GASS DR2 spectra and tables}\label{s_spectra}

\begin{figure*}[h]
\begin{center}
\includegraphics[width=15.5cm]{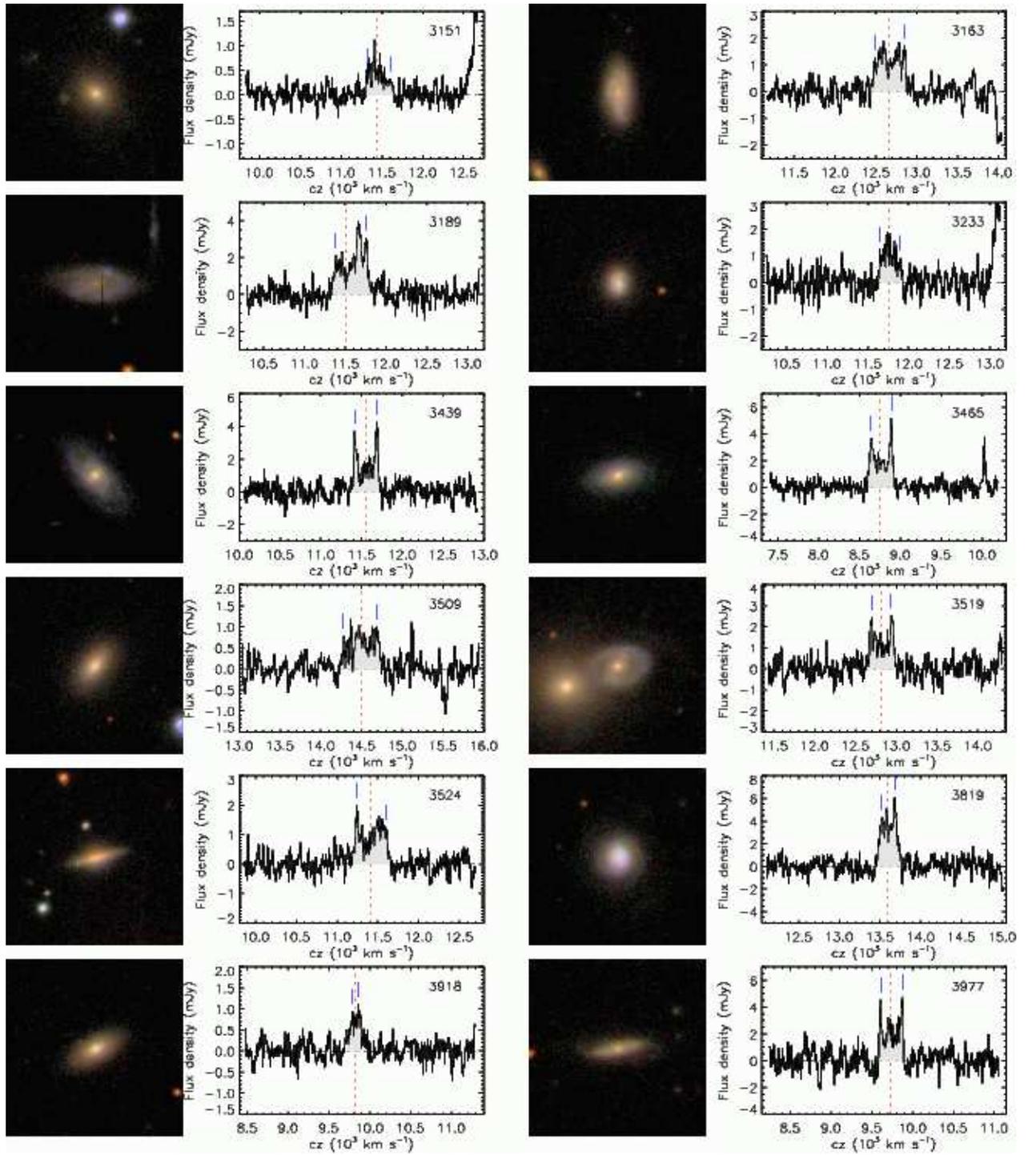}
\caption{SDSS postage stamp images (1 arcmin square) and
\hi-line profiles of the detections included in this second data
release, ordered by increasing GASS number (indicated in each spectrum). The \hi\ spectra are
calibrated, smoothed and baseline-subtracted. A dotted line and two
dashes indicate the heliocentric velocity corresponding to the SDSS
redshift and the two peaks used for width measurement, respectively.
{\em Only the first page is shown here.}
}
\label{det}
\end{center}
\end{figure*}

We present here SDSS postage stamp images, Arecibo \hi-line spectra,
and catalogs of optical, UV and \hi\ parameters for the 240 galaxies
included in this second data release.
The figures are organized as follows:
\begin{itemize}
  \item Figure~\ref{det}: \hi\ detections.
  \item Figure~\ref{dcode2}: marginal \hi\ detections (cataloged as ``quality
        code 2'' in Table~\ref{t_det}.
  \item Figure~\ref{dcode3}: marginal \hi\ detections that are also
        confused within the Arecibo beam (cataloged as ``quality
        code 3'' in Table~\ref{t_det}.
  \item Figure~\ref{dcode5}: \hi\ detections that are
        confused within the Arecibo beam (cataloged as ``quality
        code 5'' in Table~\ref{t_det}.
  \item Figure~\ref{ndet}: \hi\ non-detections.
\end{itemize}

The tables include SDSS and UV parameters for the 240 galaxies
(Table~\ref{t_sdss}), \hi\ measurements for the 133 detections
(Table~\ref{t_det}), and \hi\ upper limits for the 107 non-detections
(Table~\ref{t_ndet}). For the detailed content of the tables, see
Section~\ref{s_dr2}. Notes on individual objects (marked with an asterisk 
in the last column of Tables~\ref{t_det} and \ref{t_ndet})
are reported in Appendix~\ref{s_notes}.

\setcounter{figure}{1}
\begin{figure*}
\begin{center}
\includegraphics[width=15.5cm]{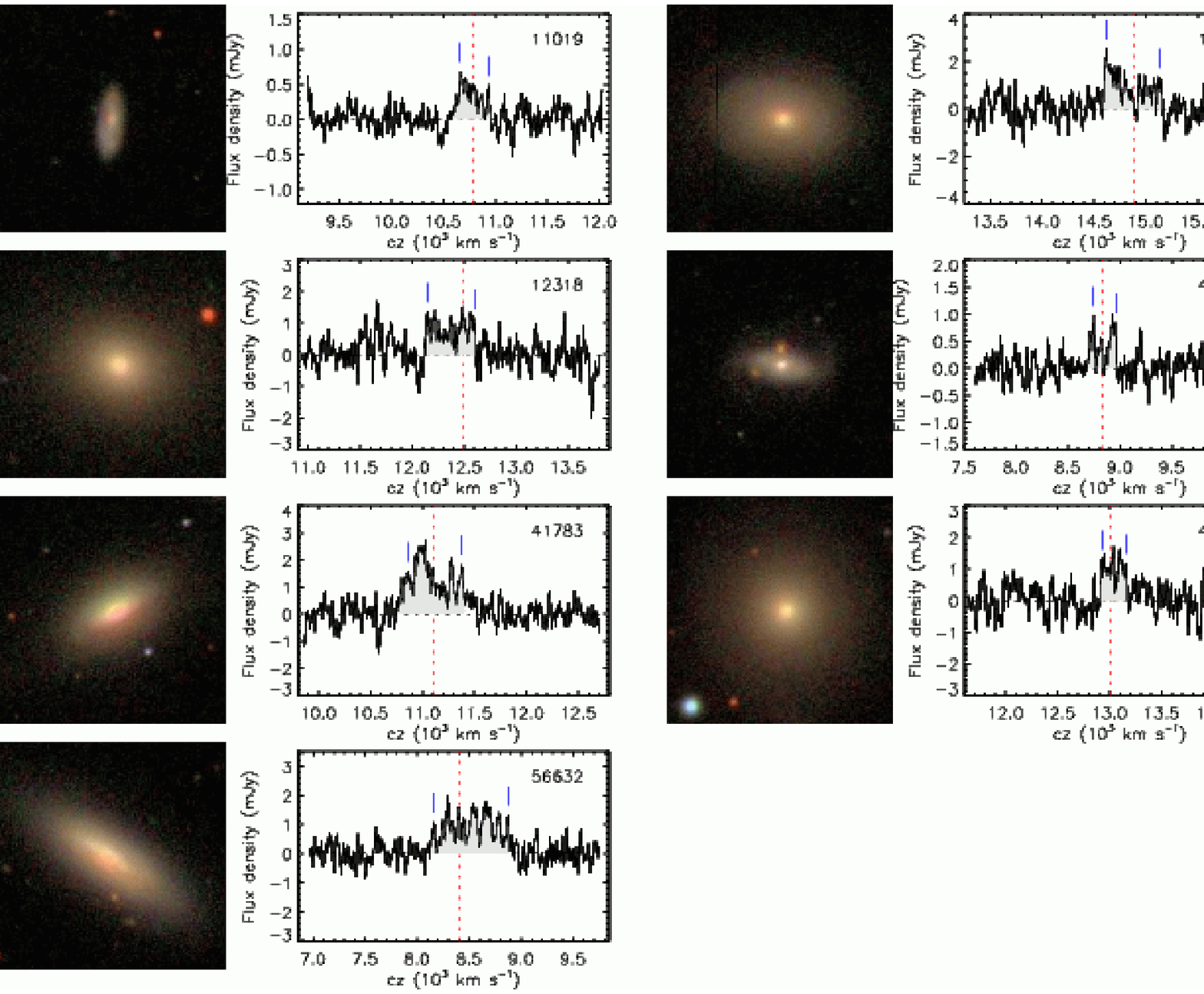}
\caption{Same as Figure~\ref{det} for marginal detections (code 2).}
\label{dcode2}
\end{center}
\end{figure*}

\begin{figure*}
\begin{center}
\includegraphics[width=15.5cm]{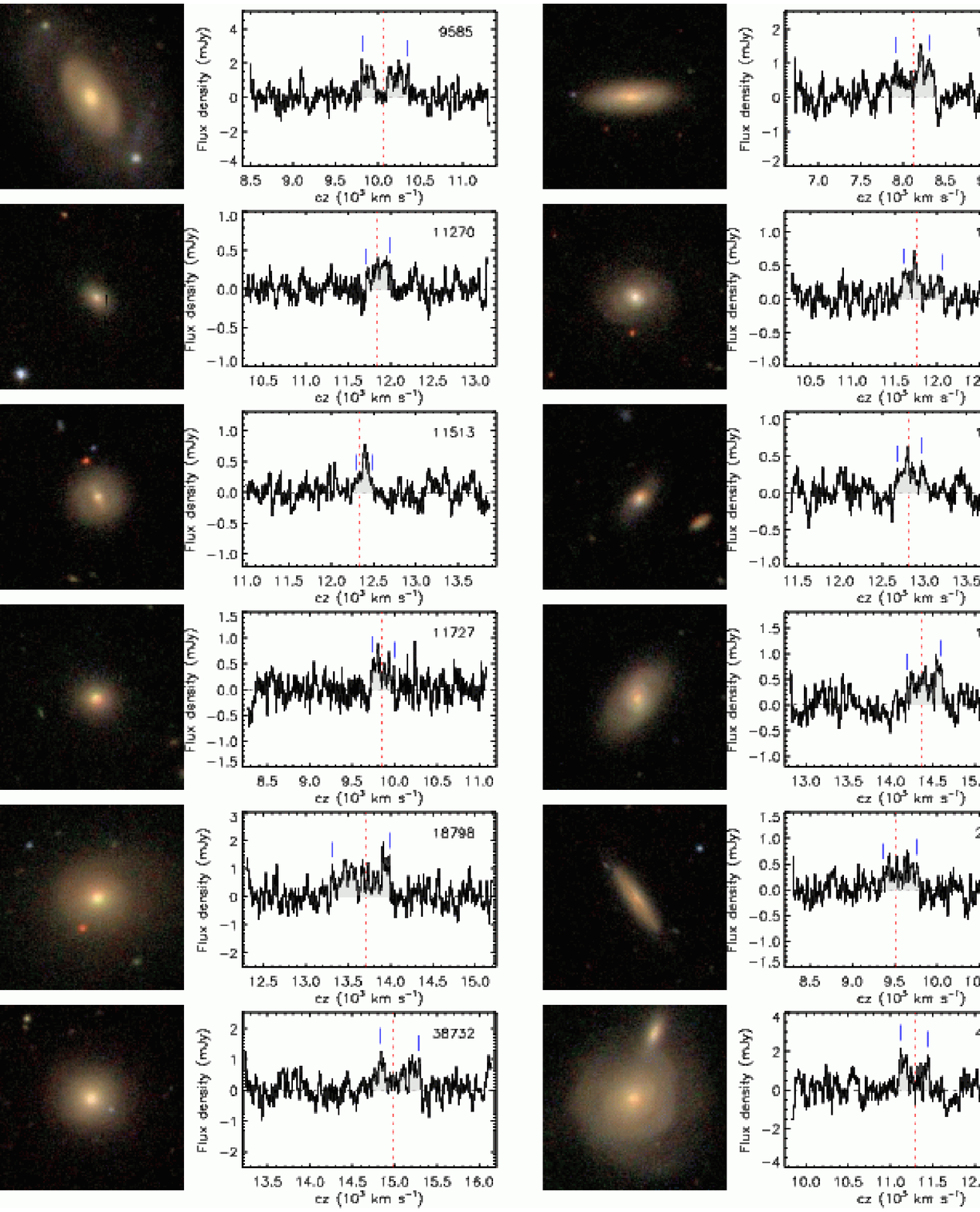}
\caption{Same as Figure~\ref{det} for marginal detections (code 3).}
\label{dcode3}
\end{center}
\end{figure*}

\begin{figure*}
\begin{center}
\includegraphics[width=15.5cm]{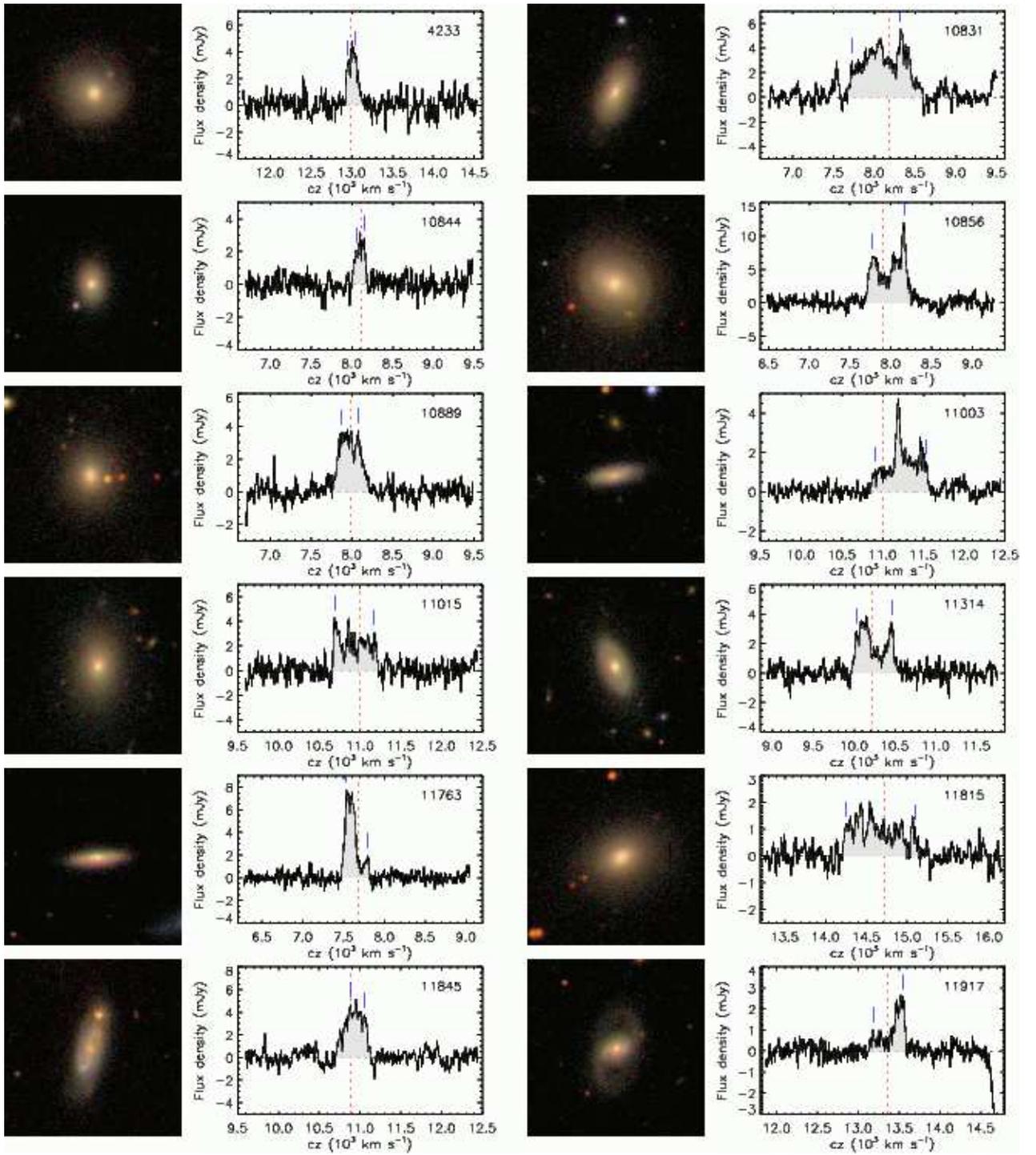}
\caption{Same as Figure~\ref{det} for confused detections (code 5).
{\em Only the first page is shown here.}}
\label{dcode5}
\end{center}
\end{figure*}

\begin{figure*}
\begin{center}
\includegraphics[width=15.5cm]{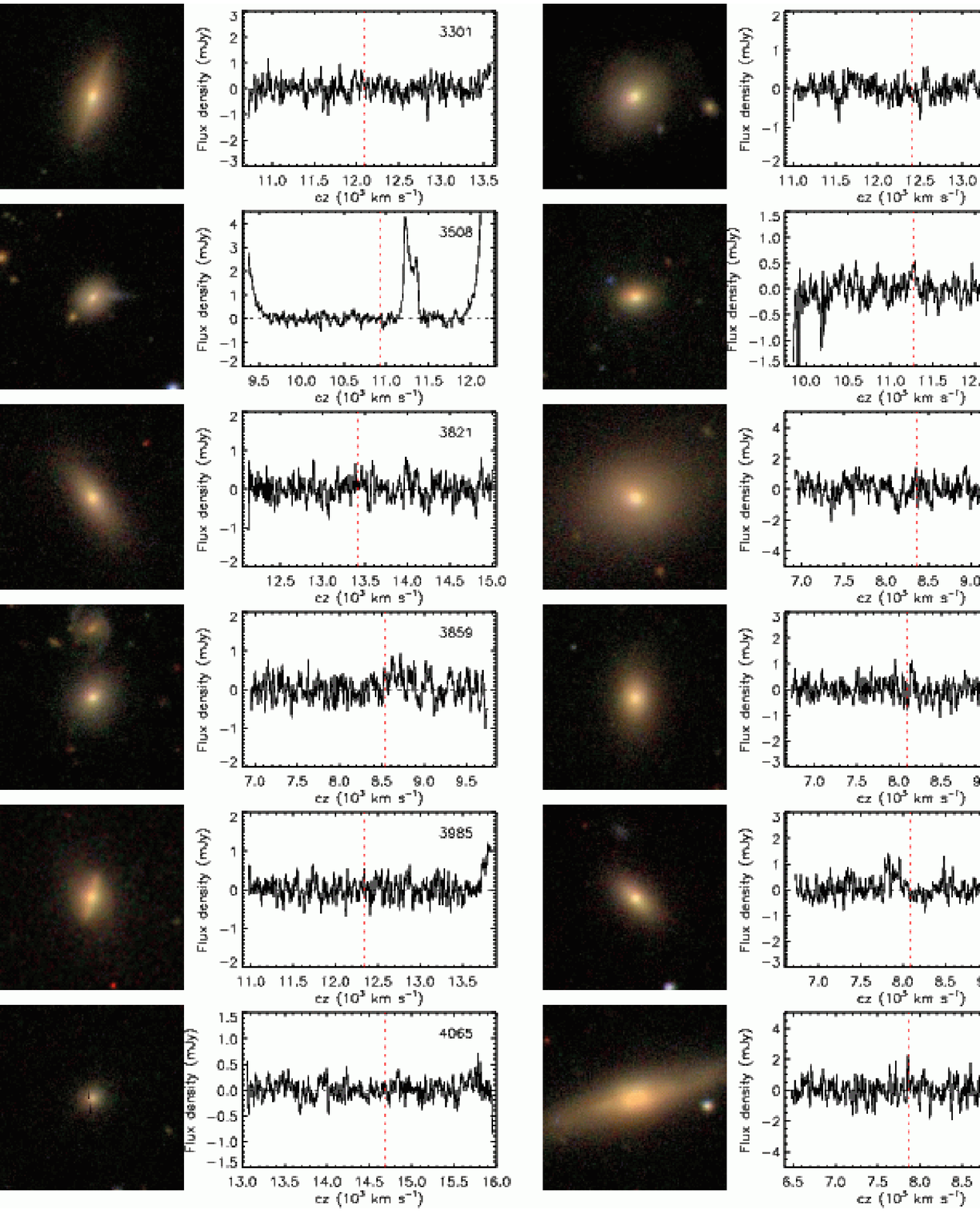}
\caption{Same as Figure~\ref{det} for non-detections.
{\em Only the first page is shown here.}
}
\label{ndet}
\end{center}
\end{figure*}

\onecolumn
\begin{landscape}
\small
\begin{longtable}{rclccccccccccccc}
\caption{\label{t_sdss}SDSS and UV Parameters of the GASS Galaxies.}\\
\hline
   &  &  &  & Log \Mst & \Rinz & $R_{50}$ & $R_{90}$ & Log \must  & ext$_r$ & r & (b/a)$_r$ & incl &\nuvr & T$_{NUV}$ & \tmax \\
GASS  & SDSS ID & Other name & $z_{\rm SDSS}$ & (\Msun) & (\arcsec) & (\arcsec)&(\arcsec) & (\Msun~kpc$^{-2}$)& (mag)& (mag) &   & (deg) & (mag)& (sec) & (min) \\
(1)  & (2)  & (3)  & (4)  & (5) & (6) & (7) & (8)  & (9)  &  (10) & (11) & (12) & (13) & (14) & (15) & (16)\\
\hline
\endfirsthead
\caption{continued.}\\
\hline
   &  &  &  & Log \Mst & \Rinz & $R_{50}$ & $R_{90}$ & Log \must  & ext$_r$ & r & (b/a)$_r$ & incl &\nuvr & T$_{NUV}$ & \tmax \\
GASS  & SDSS ID & Other name & $z_{\rm SDSS}$ & (\Msun) & (\arcsec) & (\arcsec)&(\arcsec) & (\Msun~kpc$^{-2}$)& (mag)& (mag) &   & (deg) & (mag)& (sec) & (min) \\
(1)  & (2)  & (3)  & (4)  & (5) & (6) & (7) & (8)  & (9)  &  (10) & (11) & (12) & (13) & (14) & (15) & (16)\\
\hline
\endhead
 11834 & J000047.89+141639.1 &  ...	     & 0.0362 & 10.71 &  3.21 &  3.55 & 10.51 &  9.12 &  0.17 & 15.15 & 0.341 & 74 &  3.94 &  1671 &   24 \\
 11825 & J000210.19+140810.3 &  ...	     & 0.0420 & 10.62 &  2.17 &  2.16 &  7.01 &  9.25 &  0.19 & 15.45 & 0.719 & 45 &  5.68 &   364 &   65 \\
 11917 & J000742.65+151112.7 &  ...	     & 0.0446 & 10.74 &  2.96 &  3.39 & 10.45 &  9.04 &  0.18 & 15.25 & 0.677 & 49 &  4.06 &  1680 &   48 \\
 12005 & J002420.72+144632.3 &  ...	     & 0.0309 & 10.06 &  1.71 &  1.79 &  5.09 &  9.16 &  0.17 & 16.04 & 0.348 & 73 &  5.05 &  3519 &   33 \\
 12002 & J002504.00+145815.2 &  ...	     & 0.0367 & 10.48 &  1.75 &  1.81 &  5.74 &  9.41 &  0.14 & 15.47 & 0.424 & 68 &  6.25 &  3519 &   66 \\
  3163 & J002846.83+150529.3 &  ...	     & 0.0422 & 10.63 &  3.99 &  4.23 &  9.19 &  8.72 &  0.17 & 15.12 & 0.468 & 64 &  3.82 &  3133 &   64 \\
  3151 & J003143.05+142911.2 &  ...	     & 0.0381 & 10.40 &  2.90 &  3.40 & 11.16 &  8.85 &  0.19 & 15.51 & 0.784 & 39 &  4.85 &  3133 &   77 \\
 27181 & J003248.17+144738.1 &  ...	     & 0.0418 & 10.60 &  3.38 &  3.67 & 10.23 &  8.84 &  0.18 & 15.11 & 0.781 & 40 &  3.22 &  1659 &   73 \\
\hline
\end{longtable}

\end{landscape}


\onecolumn
\begin{landscape}
\small
\begin{longtable}{rccccccccccccl}
\caption{\label{t_det}\hi\ Properties of GASS Detections.}\\
\hline
      &         &               & $T_{\rm on}$ & $\Delta v$ &     & \whi  & \whi$^c$&  $F$      &  rms & &Log \Mhi  &   & \\
GASS  & SDSS ID & $z_{\rm SDSS}$ & (min)       &  (\kms)    & $z$ &  (\kms)& (\kms) & (Jy \kms) & (mJy)& S/N  & (\Msun) & Log \Mhi/\Mst & Q \\
(1)  & (2)  & (3)  & (4)  & (5)  & (6)  & (7)  & (8)  & (9)  &  (10) & (11) & (12) & (13) & (14)\\
\hline
\endfirsthead
\caption{continued.}\\
\hline
      &         &               & $T_{\rm on}$ & $\Delta v$ &     & \whi  & \whi$^c$&  $F$      &  rms & &Log \Mhi  &   & \\
GASS  & SDSS ID & $z_{\rm SDSS}$ & (min)       &  (\kms)    & $z$ &  (\kms)& (\kms) & (Jy \kms) & (mJy)& S/N  & (\Msun) & Log \Mhi/\Mst & Q \\
(1)  & (2)  & (3)  & (4)  & (5)  & (6)  & (7)  & (8)  & (9)  &  (10) & (11) & (12) & (13) & (14)\\
\hline
\endhead
 11834 & J000047.89+141639.1 & 0.0362 &   4 & 15 & 0.036162 & 380$\pm$ 16 &  359 &  0.72$\pm$ 0.12 & 0.67 &  10.0 &  9.62 & $-$1.09 &  1*  \\
 11825 & J000210.19+140810.3 & 0.0420 &  43 & 21 & 0.042186 & 419$\pm$ 14 &  392 &  0.23$\pm$ 0.04 & 0.19 &   8.8 &  9.26 & $-$1.36 &  1   \\
 11917 & J000742.65+151112.7 & 0.0446 &  25 & 13 & 0.044594 & 401$\pm$  3 &  378 &  0.43$\pm$ 0.05 & 0.30 &  14.0 &  9.58 & $-$1.16 &  5*  \\
  3163 & J002846.83+150529.3 & 0.0422 &  15 & 21 & 0.042253 & 394$\pm$  2 &  368 &  0.48$\pm$ 0.07 & 0.32 &  11.5 &  9.58 & $-$1.05 &  1   \\
  3151 & J003143.05+142911.2 & 0.0381 &  76 & 13 & 0.038243 & 328$\pm$  9 &  309 &  0.13$\pm$ 0.02 & 0.15 &   9.5 &  8.93 & $-$1.47 &  1   \\
 27181 & J003248.17+144738.1 & 0.0418 &   5 & 21 & 0.041922 & 297$\pm$ 12 &  275 &  0.44$\pm$ 0.09 & 0.50 &   7.9 &  9.53 & $-$1.07 &  1   \\
 27167 & J003921.66+142811.5 & 0.0380 &   5 & 21 & 0.038093 & 246$\pm$ 13 &  227 &  0.62$\pm$ 0.10 & 0.56 &  10.9 &  9.60 & $-$0.77 &  1*  \\
  3189 & J004023.48+143649.4 & 0.0384 &   9 & 16 & 0.038583 & 412$\pm$ 13 &  389 &  0.83$\pm$ 0.08 & 0.44 &  16.6 &  9.74 & $-$0.31 &  1*  \\
\hline
\end{longtable}
\end{landscape}


\onecolumn
\longtab{3}{
\begin{longtable}{rccccccccccccl}
\caption{\label{t_ndet}GASS Non-detections.}\\
\hline
      &         &              &  $T_{\rm on}$ &  rms  & Log \Mhi$_{,lim}$ &  \\
GASS  & SDSS ID & $z_{\rm SDSS}$& (min)        &  (mJy)& (\Msun)  & Log \Mhi$_{,lim}$/\Mst & Note\\
(1)  & (2)  & (3)  & (4)  & (5)  & (6)  & (7) & (8)\\
\hline
\endfirsthead
\caption{continued.}\\
\hline\
      &         &              &  $T_{\rm on}$ &  rms  & Log \Mhi$_{,lim}$ &  \\
GASS  & SDSS ID & $z_{\rm SDSS}$& (min)        &  (mJy)& (\Msun)  & Log \Mhi$_{,lim}$/\Mst & Note\\
(1)  & (2)  & (3)  & (4)  & (5)  & (6)  & (7) & (8)\\
\hline
\endhead
 12005 & J002420.72+144632.3 & 0.0309 &  35 &  0.21 &  8.62 & $-$1.44 &  *  \\
 12002 & J002504.00+145815.2 & 0.0367 &  65 &  0.17 &  8.68 & $-$1.80 &  *  \\
  3301 & J005915.31+133433.3 & 0.0404 &  16 &  0.33 &  9.07 & $-$1.85 & ... \\
  3435 & J010832.07+150122.9 & 0.0414 &  33 &  0.25 &  8.97 & $-$1.79 & ... \\
  3508 & J011900.04+143643.1 & 0.0365 &  64 &  0.17 &  8.69 & $-$1.68 &  *  \\
  3757 & J012227.97+153238.4 & 0.0376 &  65 &  0.21 &  8.80 & $-$1.72 &  *  \\
  3854 & J013854.76+150117.7 & 0.0279 &   4 &  0.63 &  9.02 & $-$1.89 & ... \\
  3859 & J013901.89+145808.5 & 0.0285 &  24 &  0.26 &  8.65 & $-$1.53 &  *  \\
\hline
\end{longtable}
}



\section{Notes on individual objects}\label{s_notes}

We list here notes for galaxies marked with an asterisk in
the last column of Tables~\ref{t_det} and \ref{t_ndet}.
The galaxies are ordered by increasing GASS number. In what follows, 
AA2 is the abbreviation for ALFALFA detection code 2.\\

\noindent
{\bf Detections (Table \ref{t_det})}\\
{\bf 3189}  -- small blue smudge 20 arcsec NW, perhaps responsible for asymmetric \hi\ profile? AA2. \\
{\bf 3439}  -- AA2. \\
{\bf 3465}  -- AA2. \\
{\bf 3509}  -- RFI spikes at 1350 and 1352.2 MHz (\about 15100 and 15500 km/s); small galaxy 
 	       \about 2 arcmin S has $z=0.122$, no contamination problems. \\
{\bf 3519}  -- superposed on early type galaxy without $z$: galaxy pair or superposition effect? 
 	       Disk galaxy \about 1 arcmin E has $z=0.0597$, no contamination problems. 
 	       Stronger in polarization A but OK (polarization B has noisy baseline). \\
{\bf 3918}  -- low frequency edge uncertain, systematic error. \\
{\bf 3977}  -- AA2. \\
{\bf 3981}  -- stronger in polarization A. \\
{\bf 4030}  -- there seems to be a hint of signal from another galaxy at 1358-59 MHz (\about 13700 km/s, 
 	       no RFI there). Perhaps the small red galaxy \about 1.5 arcmin SW? There are two other galaxies, 
 	       even smaller or farther away.  \\
{\bf 4038}  -- high frequency edge uncertain, systematic error. \\
{\bf 4039}  -- blue companion \about 1.7 arcmin E ($z=0.0260$, 1384.41 MHz); another blue galaxy 
 	       \about 1.5 arcmin SW has $z=0.081$, no contamination problems; AA2. \\
{\bf 4048}  -- boards centered at wrong frequency by mistake. \\
{\bf 4057}  -- 275 mJy continuum source 1.5 arcmin away, standing waves; high frequency edge uncertain, 
 	       systematic error; AA2. \\
{\bf 4137}  -- low frequency edge uncertain, systematic error. Small galaxy \about 1 arcmin S has $z=0.155$,
 	       no contamination problems. \\
{\bf 4145}  -- notice small blue galaxy attached to GASS~4145, no redshift from SDSS, perhaps partly responsible 
 	       for asymmetric profile (which is also not well centered on SDSS redshift). Companion galaxy 
 	       \about 3 arcmin SW, SDSS J015509.55+125746.1 ($z=0.032774$, 1375.33 MHz) not detected on the side 
	       away from GASS~4145, thus unlikely to contribute to the \hi\ signal. \\
{\bf 4223}  -- several galaxies within \about 3 arcmin and with SDSS redshift, all with $z>0.08$
	       except SDSS J015717.50+144327.2 (disk galaxy \about 1.5 arcmin SE, $z=0.027101$, 1382.93 MHz) 
	       and SDSS J015705.49+144256.0 (small galaxy \about 2.5 arcmin SW, $z=0.025304$, 1385.35 MHz). 
	       In both cases, no \hi\ emission is detected in the half side that is moving away from 
	       GASS~4223, thus significant contamination is unlikely. Uncertain profile; AA2. \\
{\bf 4233}  -- confused: two blue companions with the same redshift (SDSS J015710.06+145500.9, \about 1 arcmin SE
 	       and SDSS J015708.74+145247.4, \about 3 arcmin S). \\
{\bf 5035}  -- high frequency edge uncertain, systematic error. \\
{\bf 6749}  -- small blue companion \about 40 arcsec SW, SDSS J130748.37+031057.1, $z=0.0378$ (1368.67 MHz),
 	       some contamination certain (notice that higher peak on the low velocity side is
 	       centered on the companion). \\
{\bf 7520}  -- high frequency edge uncertain, systematic error. Stronger in polarization A. Small blue 
 	       companion \about 3 arcmin N (SDSS J143318.34+024205.0, $z=0.028104$, 1381.58 MHz). \\
{\bf 9585}  -- two very small companions within \about 1.5 arcmin, one has the same $z$ (0.0336), the other has
 	       $z=0.0298$ (1379.30 MHz) and is not detected. \\
{\bf 10831} -- galaxy group: there are five galaxies within 3 arcmin E of GASS~10831,
 	       three of which with redshift $z=0.026$, one with redshift $z=0.029669$ and the largest one without
 	       redshift from SDSS. Very broad \hi\ profile is likely blend of emission of gas 
 	       from and/or between these galaxies. Uncertain profile.  \\ 
{\bf 10844} -- most \hi\ likely coming from very blue companion \about 1 arcmin NE, same $z$ (0.0270); 
 	       other galaxy \about 1 arcmin SE has $z=0.154$. \\
{\bf 10850} -- RFI spikes above 1374 MHz (\about 10000 km/s). \\
{\bf 10856} -- blend with GASS~10831 (SDSS J220322.59+123857.6, $z=0.027346$), \about 2 arcmin SW. \\
{\bf 10889} -- confused with UGC 11948? UGC 11948 is \about 3.5 arcmin W, no SDSS spectroscopy, but is in 
 	       the S05 archive ($cz= 7960$ \kms, observed flux of 2.63 Jy \kms, $W_{50}= 343$ \kms). \\
{\bf 10985} -- RFI at 1350 MHz (\about 15600 \kms). \\
{\bf 11003} -- blend with GASS~11004 (SDSS J223707.71+141355.0, $z=0.037666$, detected by ALFALFA and in
	       the S05 archive), \about 2.5 arcmin SW. Notice also blue companion \about 1 arcmin SW 
	       without SDSS redshift. \\
{\bf 11013} -- high frequency edge uncertain, systematic error. \\
{\bf 11015} -- most of the signal likely comes from SDSS J223654.54+142452.0, a large, blue disk 
 	       \about 2 arcmin W, $z=0.0365$ (1370.39 MHz). \\
{\bf 11019} -- stronger in polarization B. \\
{\bf 11055} -- high frequency edge uncertain, systematic error; clear detection but uncertain \hi\ parameters. \\
{\bf 11071} -- low frequency edge uncertain, systematic error. \\
{\bf 11120} -- high frequency edge uncertain, systematic error; stronger in polarization B. \\
{\bf 11126} -- RFI at 1358.3 MHz (\about 13700 km/s). \\
{\bf 11231} -- AA2. \\
{\bf 11250} -- AA2. \\
{\bf 11268} -- 5 other galaxies within 3 arcmin, most at slightly smaller redshift; the largest is GASS~11267,
     	       \about 1 arcmin SE ($z=0.040047$, 1365.75 MHz, i.e. 2.3 MHz higher than GASS~11268), not clearly 
 	       detected here (also targeted separately and listed as a non-detection). The bump in the baseline 
	       next to the \hi\ profile, on the low velocity side, is most likely due to these companions, but 
	       the \hi\ signal does not seem significantly affected. \\
{\bf 11270} -- low frequency edge uncertain, systematic error; much clearer in polarization B. \\
{\bf 11285} -- high frequency edge uncertain, systematic error. \\
{\bf 11303} -- low frequency edge uncertain, systematic error. \\
{\bf 11314} -- blend with GASS~11311 (SDSS J231229.22+135632.1, $z=0.034137$, ALFALFA detection), \about 
	       1.5 arcmin E. Notice also GASS~11312 (SDSS J231225.99+135450.1, $z=0.033927$), \about 2 arcmin S.
	       Low frequency edge uncertain, systematic error. \\
{\bf 11383} -- strong detection at \about 1375.5 MHz in boards 3 and 4, perhaps the blue galaxy 
 	       \about 30 arcsec SW (SDSS J232547.39+140245.2, no redshift)? The negative peak in the
	       baseline at 1374 MHz (\about 10100 km/s) is present in 4-minute pairs only, most likely
 	       due to a galaxy in the off scan (there is small blue galaxy at that position without 
	       SDSS redshift). AA2. \\
{\bf 11462} -- companion \about 2 arcmin E (SDSS J231803.74+135400.8, $z=0.040677$) possibly responsible for
 	       the enhanced peak at \about 12100 \kms. \\
{\bf 11513} -- uncertain profile, marginal detection. \\
{\bf 11514} -- uncertain profile; marginal detection, stronger in polarization B. Small galaxy 20 arcsec SW, 
 	       no optical redshift. \\
{\bf 11763} -- blend with blue companion, SDSS J234155.60+151345.9 ($z=0.025333$), \about 30 arcsec SW. \\
{\bf 11808} -- low frequency edge uncertain, systematic error. \\
{\bf 11815} -- Blue companion \about 2 arcmin SE, SDSS J235313.68+160228.3 ($z=0.048889$, 1354.20 MHz); 
               notice also blue galaxy \about 1 arcmin SE, no SDSS redshift. Confused. RFI double
	       spike at 1351 MHz (outside galaxy profile), eight channels replaced by interpolation. \\
{\bf 11824} -- RFI spike at 1374.2 MHz (\about 10000 km/s). \\
{\bf 11834} -- RFI at 1366-68 MHz (11600-11900 \kms). Blue disk \about 1 arcmin S, no redshift, 
 	       contamination possible. AA2. \\
{\bf 11845} -- RFI spike at 1375 MHz (\about 9800 km/s); blend, profile peaks uncertain.
 	       Blue companion, SDSS J235645.33+135216.4, \about 2 arcmin S with
 	       the same redshift, some contamination certain; notice that the companion is GASS~11774 
	       (AGC 331022), detected in both ALFALFA and S05 archive (observed flux of 1.96 Jy \kms, 
	       $cz=10939$ \kms, $W_{50}=323$ \kms). \\
{\bf 11917} -- blend with blue companion, SDSS J000744.11+151013.3 ($z=0.045049$), \about 1 arcmin S. \\
{\bf 12318} -- very likely blended with SDSS J111440.82+040039.8 \about 1 arcmin SW, $z=0.040675$ (1364.89 MHz); 
 	       perhaps hint of \hi\ from another companion \about 1 arcmin SE, SDSS J111447.92+040058.6, with 
	       $z=0.038769$ (1367.89 MHz, \about 11600 km/s). \\
{\bf 13158} -- profile peaks uncertain. Three small galaxies within \about 1.5 arcmin: two with redshifts close 
	       to that of GASS~13158 ($z=0.0400$ and 0.0401, which are displaced enough from the target that
 	       should be visible if detected), and one without optical $z$; AA2. \\
{\bf 15151} -- AA2. \\
{\bf 15166} -- 1.4 Jy continuum source at 6.5 arcmin and 26 mJy at 1 arcmin, standing waves, poor baseline fit.\\
{\bf 15211} -- early-type companion \about 2 arcmin E, strong contamination unlikely. \\
{\bf 18087} -- Three galaxies within 4 arcmin and with SDSS redshifts are in the background. \\
{\bf 18673} -- two small blue galaxies within 1 arcmin, no SDSS redshift; two galaxies at \about 2.5 arcmin S 
	       and SW have $z=0.08$. \\
{\bf 18681} -- detected blue companion in board 4, \about 1395.3 MHz: UGC~5308, 2 arcmin N,
 	       $z=0.017872$ from NED (1395.47 MHz). Small disk galaxy \about 1 arcmin N, SDSS J095323.73+075113.7 
	       ($z=0.039586$), unlikely to contaminate \hi\ signal. \\
{\bf 18686} -- UGC 5304, blue galaxy pair, \about 2 arcmin NE (one of the two galaxies, SDSS J095310.37+075224.8, 
    	       has $z=0.0407$, the other has no redshift in SDSS), some contamination certain. \\
{\bf 18702} -- the galaxy \about 1.5 arcmin W has $z=0.097$. A few small galaxies without SDSS redshift within 1 arcmin,
 	       one of which is likely responsible for the narrow spike within the \hi\ profile. \\
{\bf 18707} -- AA2. \\
{\bf 18798} -- high frequency edge uncertain, systematic error.
 	       Marginal detection. Small galaxy 80 arcsec E, SDSS J101334.44+075423.7, $z=0.045321$ (1358.82 MHz). 
 	       Notice asymmetric light distribution of GASS 18798 opposite to companion. Two other small galaxies 
 	       within 2 arcmin have $z=0.162$. Clearer in polarization B. \\
{\bf 18862} -- blue companion \about 1.5 arcmin S, offset by 0.8 MHz in frequency (SDSS J102241.78+081833.0, 
 	       $z=0.045199$, 1358.98 MHz), does not appear to contaminate the \hi\ signal significantly;
 	       also notice companion \about 3 arcmin N (SDSS J102236.67+082237.8, $z=0.044278$, 1360.18 MHz). \\
{\bf 20292} -- AA2. \\
{\bf 20448} -- detected also blue companion, \about 2 arcmin NW (SDSS J095808.05+110832.2, $z=0.030958$). \\
{\bf 23026} -- two companions (SDSS J102726.74+110543.9, \about 1.5 arcmin NE, $z=0.032848$
 	       and SDSS J102714.26+110340.0, \about 2 arcmin SW, $z=0.032969$), some contamination possible. 
               Detected another companion centered at $cz$ \about 8700 \kms\ and without SDSS redshift, 
	       perhaps SDSS J102715.87+110518.0 (small and blue, strong UV emission, \about 1.5 arcmin NW)? \\
{\bf 23757} -- profile edges uncertain, systematic error. \\
{\bf 24437} -- blend with blue companion, SDSS J123946.66+133159.8 ($z=0.046494$), \about 1 arcmin SW. 
               High frequency edge uncertain, systematic error. \\
{\bf 25296} -- profile peaks uncertain. \\
{\bf 26639} -- high frequency edge uncertain, systematic error. Signal present in both polarizations, but with 
 	       different shape. Blue companion \about 2.5 arcmin NE, SDSS J104055.84+134823.1 
 	       ($z=0.032261$, 1376.01 MHz), does not appear to be detected. \\
{\bf 27167} -- profile peaks uncertain. Galaxy \about 1 arcmin SE has $z=0.018$, no contamination
	       problems. \\
{\bf 28482} -- blend: blue companion \about 2.5 arcmin E, SDSS J124317.30+111241.5 
 	       ($z=0.027649$, 1382.19 MHz), and small blue satellite connected to GASS~28482 certainly
	       contributing to the \hi\ signal. \\
{\bf 29555} -- small blobs around without SDSS redshift. \\
{\bf 29892} -- blend with SDSS J112941.76+151851.3, very blue companion \about 1.5 arcmin SW ($z=0.037872$, 
	       1368.58 MHz). Notice two other galaxies at \about 3.5 arcmin with very similar redshift. AA2. \\
{\bf 30854} -- small, almost face-on blue companion \about 2.5 arcmin SW ($z=0.024564$, 1386.35 MHz) not detected.
{\bf 38538} -- RFI spike at 1372 MHz (\about 10500 \kms). \\
{\bf 38706} -- galaxy pair (UGC 9561/Arp 173), both galaxies are blue and morphologically disturbed; blend; 
	       profile edges uncertain. \\
{\bf 38732} -- high frequency edge uncertain, systematic error. Marginal detection.
 	       Companion galaxy \about 1.5 arcmin SSW, SDSS J145030.45+093626.5 ($z=0.050193$, 1352.52 MHz), has a
 	       warped disk. Some contamination almost certain. Notice another two, smaller galaxies \about 3 arcmin  
 	       from GASS 38732 with approximately the same redshift. \\
{\bf 38923} -- detected companion? The feature at 1358 MHz (\about 13800 \kms) is real, and could be part 
	       of the \hi\ profile of the blue galaxy \about 4 arcmin SE (SDSS J145806.23+105716.3, 
	       $z=0.045451$, 1358.65 MHz); there is also a small blue galaxy \about 2 arcmin SW,
	       (SDSS J145751.90+105917.8, without redshift. \\
{\bf 40260} -- RFI spike at 1372 MHz (\about 10500 \kms). Likely blend with SDSS J135836.27+132858.6, a
 	       large disk galaxy \about 2.5 arcmin E, $z=0.039954$ (1365.84 MHz; but SDSS fiber is not centered 
	       on the galaxy). Stronger in polarization A. \\
{\bf 40425} -- AA2. \\
{\bf 41303} -- confused/blend with blue companion, SDSS J140749.05+094919.2 ($z=0.037381$, 1369.22 MHz),
               a distorted spiral \about 2 arcmin E. \\
{\bf 41783} -- uncertain profile, systematic error; AA2.  \\
{\bf 41793} -- marginal detection but well centered on SDSS redshift. Possible contamination from small companion
 	       \about 2 arcmin N (SDSS J145843.76+080057.6, $z=0.037874$, 1368.57 MHz); also small galaxy 
	       \about 25 arcsec N, SDSS J145840.87+075926.9, might be merging with GASS 41793. \\
{\bf 42287} -- Small companion \about 1.5 arcmin SE (SDSS J154251.89+245826.3, $z=0.033691$), marginal 
	       contamination possible; blue galaxy \about 1.5 arcmin NW has $z=0.070$. \\
{\bf 44846} -- companion 2 arcmin N ($z=031119$) is an elliptical galaxy, contamination unlikely; blue
 	       disk \about 2.5 arcmin NE ($z=0.028982$, 1380.40 MHz) not detected. AA2. \\
{\bf 46068} -- tiny companion 2 arcmin N, SDSS J143709.19+251658.5, $z=0.034017$. \\
{\bf 51190} -- high frequency edge uncertain, systematic error. Small blue galaxy \about 1 arcmin E, no 
               SDSS redshift, perhaps responsible for this very asymmetric \hi\ profile? Three galaxies 
	       \about 2 arcmin W of GASS 51190 all have $z>0.07$. \\
{\bf 56304} -- AA2. \\
{\bf 56632} -- uncertain profile; companion galaxy \about 3 arcmin E, SDSS J090450.16+133342.8 ($z=0.028346$, 
	       1381.25 MHz), might contribute some signal. \\
{\bf 57099} -- RFI spike at 1352.2 MHz (\about 15100 \kms). Disturbed optical appearance, likely merging; 
	       blue companion \about 2.5 arcmin W, SDSS J104210.89+152136.1 ($z=0.048828$, 1354.28 MHz), 
	       unlikely to contaminate signal significantly. \\
	         
\noindent
{\bf Non-detections (Table \ref{t_ndet})}\\
{\bf 3508}  -- detected companion, SDSS J011907.98+143555.6, \about 2 arcmin E ($z=0.037748$, 1368.74 MHz). \\
{\bf 3757}  -- RFI spikes above 1373.5 MHz (below \about 10300 \kms). \\
{\bf 3859}  -- \about 20 arcsec from a spiral galaxy, no redshift in SDSS. Perhaps hint of signal centered 
	       at 1380.2 MHz. \\
{\bf 3880}  -- early type galaxy \about 1 arcmin E, similar redshift ($z=0.0259$). \\
{\bf 4040}  -- detected companion (group?), $cz$ \about 7900 \kms. There are a few small galaxies within 
               2 arcmin without SDSS redshift. Notice that NED lists a galaxy group (USGC U109) 2.7 arcmin away, 
	       whose redshift is compatible with the detection ($z=0.026055$, 1384.34 MHz). \\
{\bf 4119}  -- the feature at 13000 \kms\ is well centered on the blue companion \about 2 arcmin S 
               (SDSS J014632.27+135530.2, $z=0.043488$, 1361.21 MHz), but the galaxy is clearly detected 
	       only in polarization A. \\
{\bf 9601}  -- group: two companions with $z=0.0330$ (1375.03 MHz; 20 arcsec N, blue disk) and $z=0.0307$ 
 	       (1378.10 MHz; \about 1 arcmin SW) also not detected. Blue spiral \about 2 arcmin SE has $z=0.146$. \\
{\bf 10012} -- marginal detection? Signal most likely dominated by edge-on disk \about 2.5 arcmin E, 
	       $z=0.0282$ (1381.45 MHz; GASS~10012 has $z=0.0279$). \\
{\bf 10058} -- galaxy pair. \\
{\bf 10297} -- negative feature at 1365 MHz (\about 12100 \kms) is not RFI and is present in both polarizations. \\
{\bf 10863} -- marginal detection? Notice also blue companion \about 3 arcmin NE, SDSS J220857.85+132404.0 
 	       ($z=0.027224$, 1382.76 MHz) that might contribute to marginal signal. \\
{\bf 10879} -- RFI spike at 1375 MHz (far from galaxy), four channels replaced by interpolation. \\
{\bf 10952} -- near bright star. \\
{\bf 10979} -- detected blue companion, SDSS J221900.09+131019.1, \about 1 arcmin S ($z=0.027923$, 1381.82 MHz). \\
{\bf 11087} -- RFI spikes above 1374 MHz (below \about 10100 \kms). \\
{\bf 11267} -- companion of GASS~11268 (see detections), marginally detected here (between 12300 and 12700 km/s); 
	       four additional companions within 3 arcmin, one of which is connected to 11267 by a stellar bridge. 
	       The peak at \about 12150 \kms\ might be due to SDSS J231137.64+150443.8, a galaxy \about 2 arcmin SW 
	       with $z=0.040764$ (1364.77 MHz). \\
{\bf 11357} -- detected blue, irregular companion \about 1 arcmin NW, SDSS J232620.87+135857.0 
 	       ($z=0.043973$, 1360.58 MHz). \\
{\bf 11366} -- detected blue companion \about 1 arcmin SW, SDSS J232618.72+141150.5 
 	       ($z=0.041754$, 1363.48 MHz). \\
{\bf 11397} -- near bright star; detected blue companion \about 2.5' SE, SDSS J232453.13+142059.5 
	       ($z=0.040886$, 1364.61 MHz). \\
{\bf 11582} -- detected companion in board 4, \about 1377.5 MHz: perhaps SDSS J232751.00+142820.8, very
	       blue gal. \about 40 arcsec E without redshift? \\
{\bf 12002} -- RFI spike at 1375 MHz (\about 9900 \kms). \\
{\bf 12005} -- blue companion \about 2 arcmin SW (SDSS J002416.73+144505.9, $z=0.030938$, 1377.78 MHz)
	       also not detected. \\
{\bf 12970} -- galaxy group; GASS~12967 is 2 arcmin N, same redshift; marginally detected companion 
	       \about 1 arcmin SE, $z=0.0412$ (1364.20 MHz)? \\
{\bf 13091} -- blue galaxy \about 80 arcsec W has $z=0.040424$, well away from GASS~13091. \\
{\bf 17840} -- detected blue companion in board 4, \about 1391.5 MHz: SDSS J111250.15+093139.1, \about 1 arcmin W,
 	       $z=0.020879$ (1391.36 MHz). \\
{\bf 18482} -- detected blue companion, SDSS J123056.18+090548.9 ($z=0.038956$, 1367.15 MHz), \about 1 arcmin NE. \\
{\bf 18872} -- red companion (SDSS J102034.04+075106.5, \about 1.5 arcmin W, $z=0.044115$) also not detected. \\
{\bf 18875} -- three small companions within 2-4 arcmin also not detected. \\
{\bf 20371} -- detected blue companion \about 1.5 arcmin E (SDSS J095407.95+103625.6, $z=0.040392$). \\
{\bf 23228} -- companion \about 1.5 arcmin SE, same redshift. \\
{\bf 23437} -- detected blue, irregular companion \about 2 arcmin SW, SDSS J105144.24+115735.4 
 	       ($z=0.046049$, 1357.88 MHz), but stronger in polarization B. \\
{\bf 23563} -- small companion \about 2.5 arcmin S. \\
{\bf 24366} -- hint of detection? It would be a 3.1 sigma detection, dubious.
 	       Companion galaxy \about 1.5 arcmin NW (GASS~24364, SDSS J123936.04+122619.9, $z=0.040856$) has a
	       very disturbed morphology. Notice another galaxy \about 4 arcmin NE 
	       with similar redshift (SDSS J123951.16+122646.8, $z=0.040727$). \\
{\bf 25572} -- detected blue, asymmetric companion \about 80 arcsec E, SDSS J140705.05+130013.3 
 	       ($z=0.027756$, 1382.05 MHz). The small peak at 8000 \kms\ is present in polarization A only. \\
{\bf 25844} -- marginally detected companion, SDSS J091951.55+100812.9 \about 2.5 arcmin NW 
	       ($z=0.033534$, 1374.32 MHz). \\
{\bf 26336} -- RFI spike at 1372 MHz (\about 10500 \kms). Two companions within 3 arcmin. \\
{\bf 28461} -- galaxy pair: the companion is GASS~28462, \about 1 arcmin W, also not detected. \\
{\bf 28462} -- galaxy pair: the companion is GASS~28461, \about 1 arcmin E, also not detected. \\
{\bf 28551} -- RFI spike at 1352.2 MHz (\about 15100 \kms). Small companion \about 1 arcmin SW, same redshift. \\
{\bf 30847} -- hint of detection? It would be a 3.8 sigma detection, dubious. Large galaxy \about 1.5 arcmin NE
	       (SDSS J141838.81+072238.4, $z=0.025667$, 1384.86 MHz) not detected. \\
{\bf 38728} -- detected companion without SDSS redshift? Optical counterpart not obvious, perhaps 
	       SDSS J145052.31+094028.0 (\about 1 arcmin NW)? \\
{\bf 39120} -- hint of detection? \\
{\bf 39211} -- hint of detection? \\
{\bf 39600} -- GASS~39567, detected in DR1, is \about 2.5 arcmin S, different redshift ($z=0.031$); blue galaxy 
	       \about 2 arcmin NW is in the foreground ($z=0.006$). \\
{\bf 41444} -- marginally detected companion, a spiral galaxy \about 2 arcmin SW, SDSS J141541.05+085318.4 
	       ($z=0.02942$, 1379.81 MHz), but signal is visible in polarization B only; notice also two 
 	       early-type galaxies within 2 arcmin, no redshifts. \\
{\bf 42011} -- detected companion, perhaps SDSS J151618.38+064812.9 (\about 1 arcmin NE) or SDSS J151609.56+064741.7
 	       (small and blue, \about 1.5 arcmin W), both without SDSS redshifts. Two other galaxies within 3 arcmin
 	       have redshifts that do not match the \hi\ detection  (SDSS J151610.69+064640.8, $z=0.033492$
	       and SDSS J151618.58+064517.6, $z=0.035204$). \\
{\bf 42174} -- edge-on disk \about 2.5 arcmin NE ($z=0.033545$, 1374.30 MHz) also not detected. \\
{\bf 46564} -- it looks like a marginal detection, but the signal is in polarization A only, and is not well 
	       centered on the SDSS redshift; also galaxy has a low inclination, not consistent with wide profile. \\
{\bf 54233} -- companion 2 arcmin N. \\
{\bf 54763} -- small galaxy \about 2 arcmin SE ($z=0.045834$, 1358.16 MHz) also not detected; small galaxies without 
	       SDSS redshift nearby. \\
{\bf 54986} -- companion (SDSS J105734.98+273020.4, $z=0.04595$, 1358.01 MHz) \about 1 arcmin NW also not detected. \\
{\bf 56319} -- near bright star. \\

\end{document}